\documentclass[twocolumn]{aastex631}

\usepackage{graphicx}	% Including figure files
\usepackage{colortbl}
\usepackage{moresize}
\def\farcs{\hbox{$.\!\!^{\prime\prime}$}} 
\shorttitle{10\,GHz Survey of the GOODS-N field}
\shortauthors{Jiménez-Andrade et al.}
\begin{document}

\title{A Census of the Deep Radio Sky with the VLA I:\\  10\,GHz Survey of the GOODS-N field
\footnote{\url{https://science.nrao.edu/science/surveys/vla-x-gn/home}}}

\author[0000-0002-0786-7307]{Eric F. Jiménez-Andrade}
\affiliation{Instituto de Radioastronomía y Astrofísica, Universidad Nacional Autónoma de México, Antigua Carretera a Pátzcuaro \# 8701,\\ Ex-Hda. San José de la Huerta, Morelia, Michoacán, México C.P. 58089}
\affiliation{National Radio Astronomy Observatory, 520 Edgemont Road, Charlottesville, VA 22903, USA}

\author[0000-0001-7089-7325]{Eric J. Murphy}
\affiliation{National Radio Astronomy Observatory, 520 Edgemont Road, Charlottesville, VA 22903, USA}

%\collaboration{6}{(AAS Journals Data Editors)}

\author[0000-0003-3168-5922]{Emmanuel Momjian}
\affiliation{National Radio Astronomy Observatory, P.O Box O, Socorro, NM 87801, USA}

\author[0000-0001-7583-0621]{James J. Condon}
\affiliation{Unaffiliated}

\author[0000-0003-4724-1939]{Ranga-Ram Chary}
\affiliation{ IPAC, California Institute of Technology, MC 314-6, 1200 E. California Boulevard, Pasadena, CA 91125, USA}

\author[0000-0001-9885-0676]{Russ Taylor}
\affiliation{Inter-University Institute for Data Intensive Astronomy, University of Cape Town, Rondebosch, 7701, South Africa}
\affiliation{Department of Astronomy, University of Cape Town, Rondebosch, Cape Town, 7701, South Africa}
\affiliation{Department of Physics and Astronomy, University of the Western Cape, Bellville, Cape Town, 7535, South Africa}

\author[0000-0001-5414-5131]{Mark Dickinson}
\affiliation{National Optical Astronomy Observatories, 950 N Cherry Avenue, Tucson, AZ 85719, USA}

% \author{Julie Steffen}
% \affiliation{AAS Director of Publishing}
% \affiliation{American Astronomical Society \\
% 1667 K Street NW, Suite 800 \\
% Washington, DC 20006, USA}

% \author{Magaret Donnelly}
% \affiliation{IOP Publishing, Washington, DC 20005}



\begin{abstract}

We present the first high-resolution, high-frequency radio continuum survey  that fully maps an extragalactic deep field: the 10\,GHz survey of the Great Observatories Origins Deep Survey-North (GOODS-N) field. This is a Large Program of the  {\it Karl G. Jansky}  Very Large Array that allocated 380 hours of  observations using the X-band ($8-12$\,GHz) receivers, leading to a 10\,GHz mosaic of the GOODS-field   with an average rms noise  $\sigma_{\rm n}=671\,\rm nJy\,beam^{-1}$ and angular resolution  $\theta_{1/2}=0\farcs22$ across 297\,$\rm arcmin^2$. To maximize the  brightness sensitivity we also produce a low-resolution mosaic with  $\theta_{1/2}=1\farcs0$ and $\sigma_{\rm n}=968\,\rm nJy\,beam^{-1}$, from which we derive our master catalog containing 256 radio sources detected with peak signal-to-noise ratio  $\geq 5$. Radio source size and flux density estimates from the high-resolution mosaic are provided in the master catalog as well. The total fraction of spurious sources in the catalog is 0.75\%. Monte Carlo simulations are performed to derive completeness corrections of the catalog. We find that the 10\,GHz radio source counts in the GOODS-N field agree, in general, with predictions from numerical simulations/models and expectations from 1.4 and 3\,GHz radio counts.  

\end{abstract}



% =================================================== 
\section{Introduction} \label{sec:intro}

\begin{figure*}
\centering
  \includegraphics[width=1.0\textwidth]{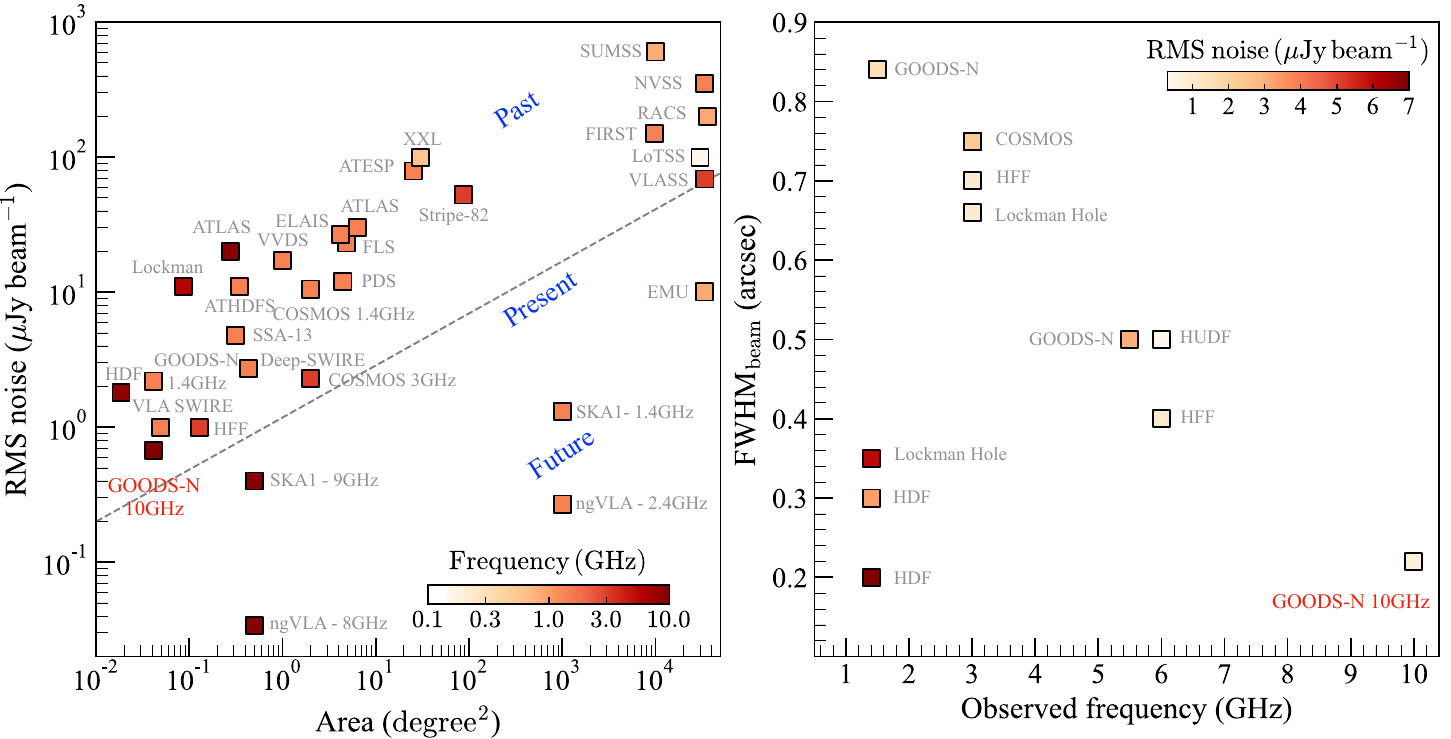}
\caption{{\it Left:} A summary of past, present, and future extragalactic radio surveys in the sensitivity {\it vs} survey area plane. Figure adapted from \citet{smolcic17}. { The data are color-coded according to the observed frequency.} The Square Kilometer Array Phase 1 (SKA1) data points are taken from the SKA1 Science Requirements. These predictions are made for a 10,000\,hours survey across 1,000\,degree$^2$  at 1.4\,GHz and a 1,000\,hours survey across 0.5\,degree$^2$ at 9\,GHz. The next generation Very Large Array (ngVLA) data points are derived using the ngVLA exposure calculator tool. We adopt the central frequencies of the Band 1 (Band 2) of the ngVLA \citep[see][for a review]{murphy22}, and assume 1,0000  (1,000)\,hours of telescope time to map a  1,000 (0.5)\,degree$^2$ region. For both telescopes,  spatial resolutions of $0\farcs5$ and $0\farcs05$ are assumed when observing at 1.4/2.4\,GHz and 8/9\,GHz, respectively.    {\it Right:}  A compilation of deep extragalactic radio maps obtained with sub-arcsec resolution \citep {muxlow05, bigss08, miettinen15, lindroos16,    rujopakarn16,  guidetti17, bondi18, cotton18,  muxlow20, jimenezandrade19, jimenezandrade21}, { color-coded according to the RMS noise},  which have been preferentially obtained at $1-3$\,GHz. The 10\,GHz survey of the GOODS-N field stands as one of the deepest ever obtained. It is also the first observational campaign that fully maps an entire extragalactic field at high frequencies, providing  some of the highest angular resolutions ever achieved in deep radio maps. 
\label{fig:surveys}}
\end{figure*}

Extragalactic surveys are essential tools to carry out statistical analysis of galaxy populations  and investigate the physical processes regulating galaxy evolution throughout cosmic time.  Radio continuum surveys at $\sim$GHz frequencies are particularly valuable, because they allow us to probe non-thermal processes like synchrotron emission  from supernova remnants \citep[e.g.,][]{dubner15}
 and  relativistic jets powered by accreting supermassive black holes \citep[e.g.,][]{miley80}. Further, thermal (free-free) radiation from H{\sc ii} regions  is detectable in the radio regime and dominates the total radio emission of star-forming galaxies (SFGs)  at frequencies { $30\,\rm GHz\lesssim \nu \lesssim 100\,\rm GHz$} \citep[e.g.,][]{murphy18, klein18}. Radio continuum surveys, therefore, provide a unique window into the  SFGs and Active Galactic Nuclei (AGN) populations \citep[see][for a review]{condon92, tadhunter16}. This has motivated the implementation of increasingly wider and deeper extragalactic radio surveys during the past  decades \citep[see left panel of Figure\,\ref{fig:surveys}, and][for a review]{simpson17}. Because the primary beam areas $(\Omega_{\rm beam}$) in radio observations are inversely proportional to the observed frequency $(\Omega_{\rm beam} \propto 
\nu^{-2}$), most surveys of the extragalactic radio sky have been obtained  at $\approx 1-3$\,GHz \citep[e.g.,][]{afonso01, seymour04, schinnerer07, ibar09, white12,  smolcic17,  heywood21,  best23, hale21, hale23}. Moreover, galaxies are easier to detect at low frequencies (i.e., $\sim$GHz) where synchrotron-dominated emission leads to a steep spectral index $\alpha\sim -0.7$, which generally describes the radio spectral energy distribution (SED) of SFGs and AGN following $S \propto \nu^{\alpha}$ \citep[e.g.,][]{ tabatabaei17, tisanic20, An24}.

Enabled by the improved broadband and wide-field imaging capabilities  of modern radio interferometers like the    {\it Karl G. Jansky} Very Large Array (VLA) and MeerKAT, it is now possible to explore the $\mu\rm Jy$ radio source population at $\sim$GHz frequencies across $\gtrsim \rm  1\,deg^{2}$ regions
    \citep[e.g.,][]{smolcic17, matthews21, hale23} and, thereby, carry out systematic studies of radio-selected SFGs and AGNs out to $z\approx5$ \citep[e.g.,][]{smolcic17b, delvecchio17, novak17,  vardoulaki19, leslie20, matthews21b, amarantidis23}. Due to the large areal coverage of current $\sim$GHz deep radio surveys, constraints on the sub-mJy radio source counts are less influenced by sample/cosmic variance, which is essential for testing and refining theoretical models of galaxy evolution \citep[e.g.,][ and references therein]{mancuso_15, mancuso17, bonaldi19}. Furthermore, a small fraction of $\sim$GHz  radio surveys have even reached  sub-arcsec angular resolutions \citep[right panel of Figure \ref{fig:surveys};  e.g.,][]{smolcic17, muxlow20}, allowing us to investigate the radio morphological properties of high-redshift compact sources \citep[e.g.,][]{bondi18, cotton18, jimenezandrade19,  jimenezandrade21, vardoulaki19}.

    Despite the rapidly growing number of extragalactic radio continuum surveys, the high-frequency ($\sim 10-100\,\rm GHz$) radio sky has been  sparsely explored (right panel of Figure \ref{fig:surveys}). High-frequency radio surveys are important to investigate, for example:
    \begin{itemize}
        \item mechanisms for cosmic-ray energy losses, 
        \item young radio sources whose radio spectra peak at progressively higher frequencies, 
        \item corrections for astrophysical foregrounds in Cosmic Microwave Background (CMB) maps \citep[e.g.,][]{dezotti05}, 
        \item and  anomalous microwave emission \citep[AME; e.g.,][]{murphy18_ames} arising from  spinning and magnetized ultra-small dust grains. 
 \end{itemize}
Most important for studies on star formation, and the scope of this manuscript,  high-frequency observations are sensitive to free-free emission that is a better  dust-unbiased  tracer of “current” star formation \citep[e.g.,][]{murphy11}, as opposed to synchrotron that traces cumulative history of star formation. These science topics have motivated high-frequency extragalactic surveys  at $\approx 10-20$\,GHz \citep{bolton04, sadler06, whittam16, huynh19} and even 95\,GHz \citep{sadler08, gonzalez-lopez19}; nevertheless, { most of the these surveys} reached  depths $\gtrsim 0.1\,\rm mJy$ where the dominant radio source population are  AGNs. In a pioneering effort to probe the   high-frequency extragalactic sky at $\mu$Jy levels, \citet{richards98} and \citet{fomalont02} carried out single-pointing VLA observations at $\approx 8.5$\,GHz that reached up to a $\approx 1.5 \,\rm \mu Jy\,beam^{-1}$ sensitivity and synthesized beam  with FWHM of $3\farcs5$, which helped demonstrating that there is an increasing contribution from SFGs to the total radio source population in the $\mu$Jy regime. More recently, \citet{algera21} and \citet{vandervlugt21} obtained single-pointing VLA  continuum observations at 34\,GHz and 10\,GHz down to a rms noise of $1.3  \,\rm \mu Jy\,beam^{-1}$ and  $0.41\,\rm \mu Jy\,beam^{-1}$, respectively, and angular resolutions $\gtrsim 2\farcs0$. These ultra-deep observations allowed \citet{algera21} to verify the robustness of free–free emission as a SFR indicator at high redshift; consequently, high-frequency radio emission has  been used to derive the first constraints on the cosmic star formation history from free–free radio emission, which agrees with the ones inferred from other widely tested SFR indicators \citep{algera_22}. Moreover, the aforementioned 10 and 34\,GHz VLA observations led to some of the first constraints on the radio source counts in the $\mu \rm Jy$ regime at high-frequencies \citep{algera21, vandervlugt21}, albeit such results are potentially affected by sample and/or cosmic variance due to the small areal coverage of these single-pointing VLA observations.  

    \subsection{A VLA 10\,GHz  Large Program in  GOODS-N}
    
    While deep, high-frequency radio observations are becoming increasingly available, to date, these are limited to single-pointing maps with coarse angular  resolutions.   To demonstrate the feasibility of a large survey of the high-frequency radio sky with the VLA, \cite{murphy17} carried out a  pilot program using a single-pointing in the Great Observatories Origins Deep Survey-North \citep[GOODS-N;][]{dickinson03, giavalisco04} at 10\,GHz. Observing at this frequency has the distinct advantage of  yielding sub-arcsec angular resolution imaging while probing higher rest-frame frequencies of galaxies with increasing redshift, where emission becomes dominated by thermal (free-free) radiation and directly provides a dust-unbiased measurement of massive star formation activity. By targeting the GOODS-N field one also maximizes the impact of galaxy formation and evolution research, as this is one of the best-studied extragalactic fields at optical/near-infrared wavelengths. Ancillary data available in this field include extremely deep observations from the {\it Hubble Space Telescope (HST)},  {\it James Webb Space Telescope (JWST)}, {\it Spitzer Space Telescope},  {\it Chandra X-ray Observatory}, {\it Herschel Space Observatory}, and the {\it XMM-Newton Observatory} \citep[see][and references therein]{ barro19,  eisenstein23, oesch23}. There is deep, high-resolution  radio imaging in this field at 1.5\,GHz \citep{morrison10, owen18, muxlow20} and 3\,GHz 
(Jiménez-Andrade, et al. in prep).  Additional, yet shallower, 5, 5.5, and 8\,GHz data are also available for a small fraction of the GOODS-N field \citep{richards99, guidetti17, gim19}. 

    The pilot program of   \cite{murphy17} proved that combining multi-configuration VLA 10\,GHz data  significantly improves the capability to recover integrated flux densities of both extended and compact sources, measure source sizes, and obtain radio spectral indices and thermal fractions using the existing radio imaging in GOODS-N. Combining information from 10\,GHz images with circular synthesized beams with FWHM of $1\farcs0$ and $0\farcs22$ and rms noises of $1.1\rm \, \mu Jy\, beam^{-1}$ and  $572\rm\,  nJy\, beam^{-1}$, respectively,  \cite{murphy17} report the detection of  38 radio sources (above the 3.5$\sigma$ level) with an optical and/or near-infrared   counterpart  with a median redshift of $1.24\pm0.25$. The resolution of $0\farcs22$ sufficed to derive the deconvolved FHWM of all the 32 radio sources detected in the high-resolution map, leading to a median effective radius of $69\pm 13$\,mas that translates into  $\approx 509\pm114$\,pc at the median redshift of this galaxy sample.  { These radio sizes are a factor $\sim 7$ smaller, on average, than the optical size, suggesting that star formation is centrally concentrated in these 10\,GHz-detected galaxies at $z\approx 1.24$. }

Motivated by the results from the pilot program reported in \cite{murphy17}, we have conducted a VLA Large Program to produce a deep, high-resolution mosaic of the entire  GOODS-N field at 10\,GHz. The new set of observations consists of 17 VLA pointings with an angular resolution and sensitivity similar to that obtained in our single-pointing pilot program \citep{murphy17}. As a result, this is the first observational campaign that fully maps an entire extragalactic field at high frequencies and high angular resolutions reaching sub-$\mu$Jy sensitivities. { Specifically, a deep 10\,GHz mosaic with an angular resolution of $0\farcs22$ is necessary to probe the spatial distribution of massive, dust-obscured star formation in galaxies at $ 0.5\lesssim z \lesssim 4$. Measuring the structure
in the radio regime relative to the optical/ultraviolet, for example, will be key to link the level and nature of star formation and AGN activity to the stellar mass buildup in galaxies.} 

Here, we report the  radio continuum data products (mosaics and radio source catalogs) and inferred 10\,GHz radio source counts. This is the first of a series of manuscripts that will explore the radio source populations in the GOODS-N field using the 10\,GHz data reported here and recently obtained deep, high-resolution 3\,GHz data (Jiménez-Andrade, et al. in prep.).

This manuscript is organized as follows. In Section\,\ref{sec:data}, we describe the 10\,GHz VLA data set and the imaging procedure.  Section\,\ref{sec:cataloging} reports the source extraction and properties of the radio source catalogs. We assess the reliability of the radio source catalog in  Section\,\ref{sec:corrections}, while the  inferred 10\,GHz radio source counts are presented in Section\,\ref{sec:counts}. A summary and conclusions from this work are given in Section\,\ref{sec:conclusions}.

\section{Observations, data reduction, and imaging} \label{sec:data}

\subsection{Very Large Array Observations}
A total of 380 hours of observations were taken from September 2016 to March 2018 with the VLA towards the GOODS-N field using the X-band ($8-12$\,GHz) receivers (Project code: VLA 16B-320; Principal Investigator: Eric J. Murphy). The data cover a bandwidth of 4096 MHz, separated into 32  128 MHz-wide spectral windows (SPWs), and are centered at 10\,GHz. The observations were obtained with a 3s signal-averaging time and full polarisation mode, albeit this manuscript only reports the total intensity mosaic and associated radio source catalog.

300 hours of observations were taken in the A-configuration (with a maximum baseline  $B_{\rm max}=36.4$\,km) of the VLA to provide the best possible angular resolution to resolve the radio emission of high-redshift SFGs. These data are complemented with 65 and 15\,hours of observations in the B and C configuration (with $B_{\rm max}=11.1$ and 3.4\,km), respectively, to improve our sensitivity to low surface brightness structures.

\begin{figure}
\centering
  \includegraphics[width=1.0\columnwidth]{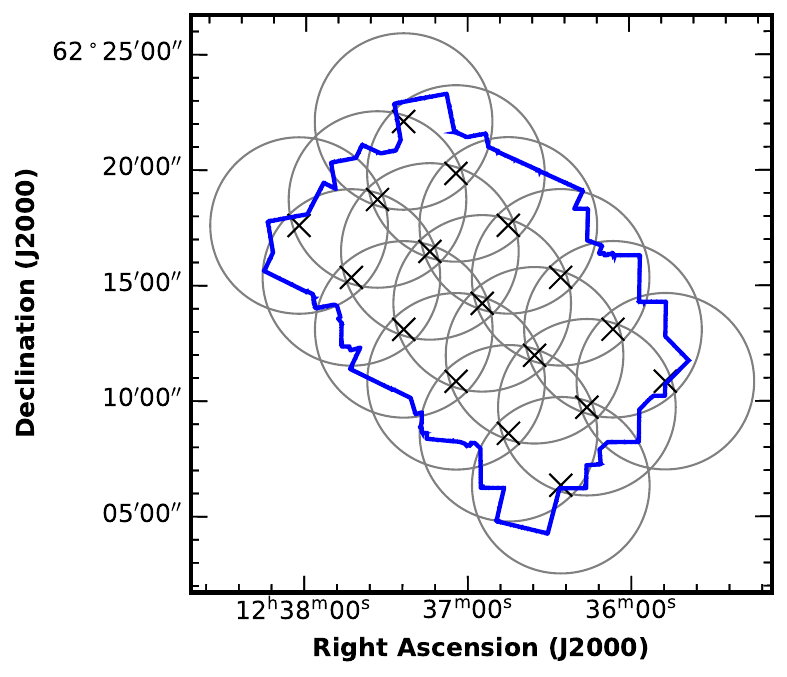}
\caption{Pointing pattern used to obtain the    10\,GHz mosaic of the GOODS-N field. The pointing centers are shown by the crosses. The circles extend out to a radius of $3\farcm82$ where the VLA primary beam response is 0.1. The blue outline  illustrates the footprint of the {\it HST}/WFC3 F160W imaging from the CANDELS program \citep{grogin11, koekemoer11}.  
\label{fig:pointing_distribution}}
\end{figure}

To obtain a  nearly uniform sensitivity across the GOODS-N field, seventeen pointings were chosen to obtain a hexagonal-pattern mosaic  (see Figure\,\ref{fig:pointing_distribution}). The separation between the pointings center is $\rm HPBW/\sqrt{2}=3\farcm18$, where the  half-power beam width (HPBW) of the VLA at 10\,GHz is $\approx4\farcm5$. The seventeen pointings were observed during each of the 5\,hours-long observing runs used during the observations.  
At the beginning of each run, 3C\,286 was observed during $\approx$15\,mins for flux density scale, polarization angle, and bandpass calibration. Then, J1302+5748 was observed for gain and phase calibration during $\approx 1.5$\,min every $\approx15$\,min when using the A and B configuration,  or every $\approx30$\,min  when observing with the C configuration. Each  pointing was visited once during the observing run and observed for $\approx15-20$\,min.

\subsection{Data Calibration}
We used the VLA calibration pipeline (version 5.6.2-3), implemented in the Common Astronomy Software Applications \citep[CASA;][]{mcmullin07, casateam22} package, to process the 76 scheduling blocks from our data set and obtain calibrated measurement sets (MSs). This pipeline is optimized to work for Stokes I continuum data by performing basic flagging (e.g., shadowed data, edge channels of sub-bands, radio frequency interference) and  deriving/applying delay, bandpass, and  gain/phase calibrations. We used the pipeline ``weblog" to verify the quality assurance (QA) of each flagging and calibration step for all the calibrated MSs. Additionally,  to further evaluate the pipeline results, we imaged the 17 pointings per each MS and inspected the resulting 1292 images. This QA process led to the identification of defective scans arising from  bad weather conditions (i.e., high phase rms values from the Atmospheric Phase Interferometer), low elevations, and high wind speeds during the observations taken in the A configuration. These defective scans, that  correspond to only 3.2\% of the total data, were  discarded outright. We also inspected the ``amplitude {\it vs} frequency'' diagnostic plots in the ``weblog" and found a satisfactory performance of the pipeline in flagging radio frequency interference (RFI). We verified that additional flagging of RFI remaining from the pipeline has a negligible impact on the imaging quality. Finally, we split the 76 calibrated MSs into 17 MSs containing  all the data from the pointings/fields (i.e., A, B, and C configuration observations) used to cover the full GOODS-N field.

\begin{figure*}
\centering
\includegraphics[width=1\textwidth]{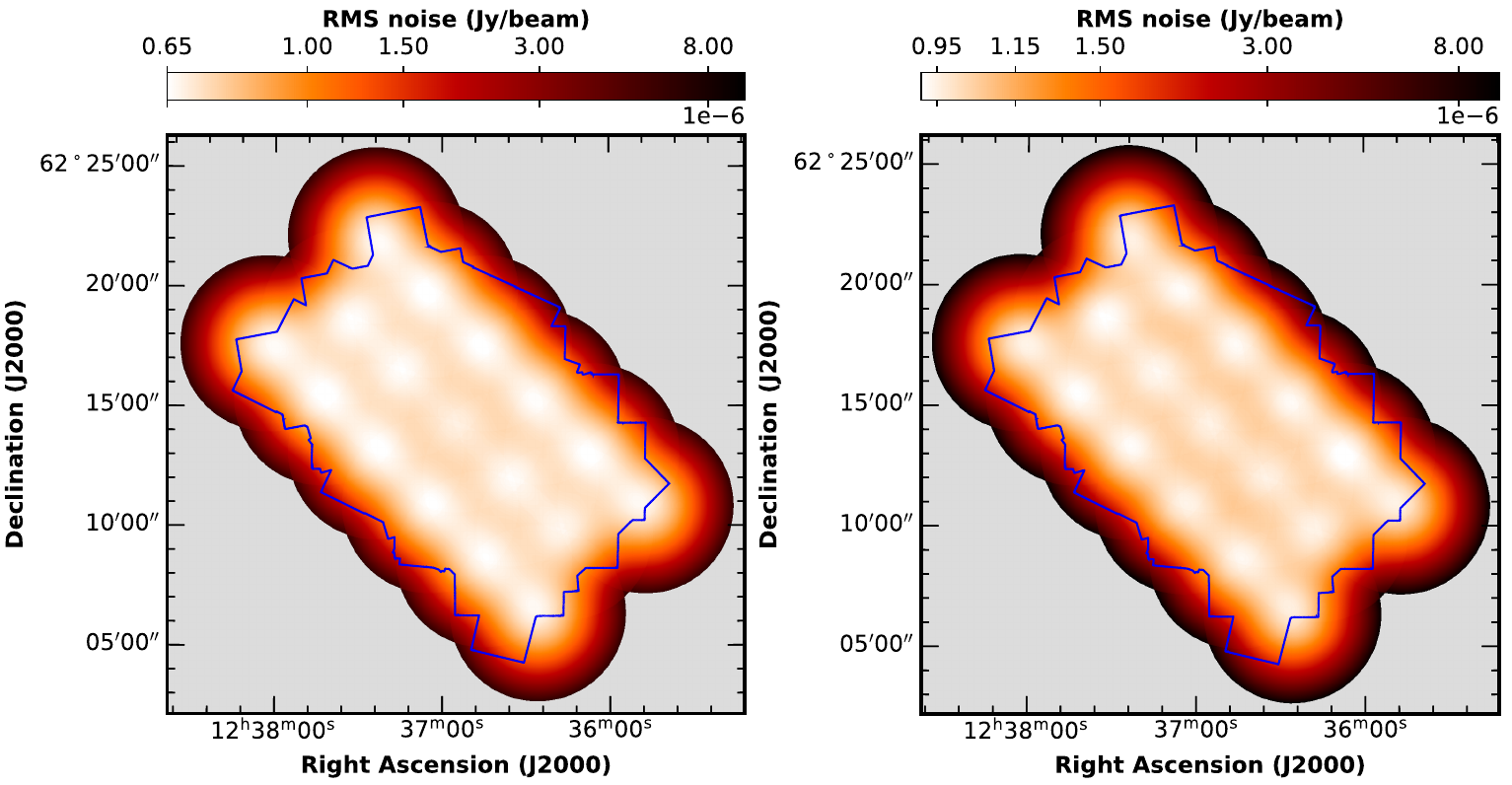}
\caption{RMS noise across the 10\,GHz mosaic of GOODS-N at $0\farcs22$ ({\it left}) and $1\farcs0$  ({\it right}) resolution. The blue outline  illustrates the footprint of the {\it HST}/WFC3 F160W imaging from the CANDELS program \citep{grogin11, koekemoer11}. Due to the pointing layout,  the rms noise variations among the pointing centers are less than 10\%.  
}
\label{fig:rms_distribution}
\end{figure*}

\subsection{Imaging}
The  calibrated MSs with the A, B, and C configuration data from the 17 fields chosen to cover the GOODS-N field were imaged with {\tt tclean} in CASA. We adopted the Multi-Term Multi Frequency Synthesis (MTMFS) imaging mode \citep{rau-cornwell11} that performs multi-scale and multi-term cleaning for wideband imaging. We set the number of Taylor coefficients used in the spectral model to {\tt nterms=2}, i.e., the spectrum is considered as a straight line with a slope, to take into account variations of the spectral structure across the image. To reconstruct the emission of complex, extended radio sources through the multi-scale cleaning implemented in MTMFS, we look for scales extending up to 16 times the FWHM of the synthesized beam. In addition, we implement the {W-projection} algorithm that corrects for a non-zero {\it w}-term arising from the sky curvature and non-coplanar baselines in wide-field imaging. In practice, this algorithm hinders the presence of artifacts around sources away from the phase center. After extensive testing with several values for the number of {W-projection} planes to use, we find that {\tt wprojplanes=64} leads to adequate imaging quality out to the regions where the primary beam response drops to $10\%$. Self-calibration was not implemented due to the faint nature of the radio sources.

\subsubsection{High-Resolution Mosaic}
\label{subsubsec:high-res_mosaic}
Following the tests performed for the pilot survey \citep{murphy17}, we adopt the Briggs weighting with {\tt robust=0.5}. The native, synthesized beam of the combined A, B, and C configuration observations is fairly Gaussian with major and minor FWHMs $\theta_{\rm maj}=0\farcs 23\times \theta_{\rm min}=0\farcs20$. For simplicity, we specified a circular Gaussian restoring beam with $\rm FWHM=0\farcs22$, as in our pilot survey \citep{murphy17}, to image each pointing individually out to a primary beam response of $10\%$. Major cycles of {\tt tclean} are run in parallel with the option {\tt parallel=True} and cleaning stops once the residuals are four times the rms noise.  The final 17 images have $10,000 \times 10,000$ pixels with a pixel scale of $0\farcs05$, covering a  $8\farcm33\times8\farcm33$ region. Following the imaging of the 17 VLA pointings towards the GOODS-N field, we used the task {\tt widebandpbcor} to  perform a wideband primary-beam correction. Then, the resulting 17 images are combined in a weighted fashion with the task {\tt linearmosaic} to obtain a linear mosaic, $I^{lm}(x)$, given by 

\begin{equation}
    I^{lm}(x) = \frac{\sum_p A(x-x_p) I_p(x) }{\sum_p A^2(x-x_p) },
    \label{eq:mosaic}
\end{equation}
where $A$ is the VLA primary beam at 10\,GHz, $I_p$ is $p^{th}$ deconvolved image, and $x_p$ the  pointing center. The resulting mosaic covers a total area of  $\rm 297\,arcmin^2$ and is centered at J2000 right ascension (RA) $12^{\rm h} 36^{\rm m} 55^{\rm s}$ and declination (DEC) $+62^\circ 14' 15'' $.   \\

\begin{figure}
\begin{center}
\includegraphics[width=1\columnwidth]{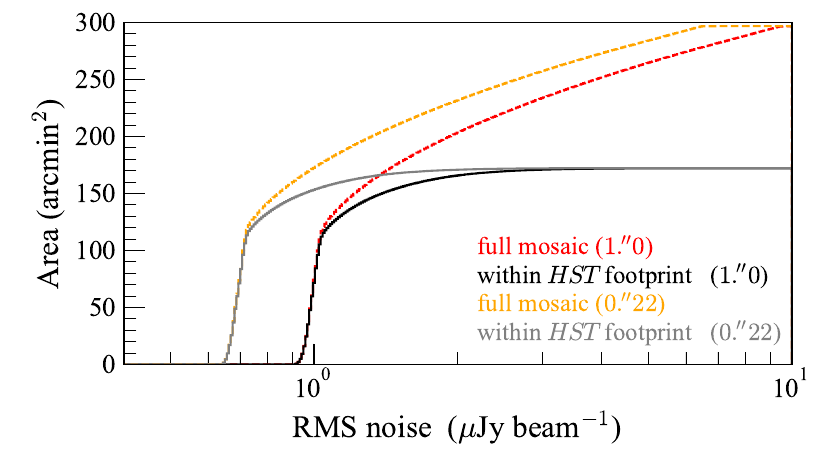}
\caption{Cumulative distribution of the total area covered down to a given rms noise  level. The black and gray lines correspond to the distribution of the rms noise  level at $0\farcs22$ and $1\farcs0$ resolution, respectively, within the footprint of the {\it HST}/WFC3 F160W imaging (covering 170\,arcmin$^2$). The orange and red lines show the distribution of the rms noise   at $0\farcs22$ and $1\farcs0$ resolution, respectively, across the entire 10\,GHz mosaic of GOODS-N (covering 297\,arcmin$^2$). Since our VLA data extend beyond the area covered by existing {\it HST} images,  the rms noise  remains nearly homogeneous within the central $\approx$120\,arcmin$^2$.   }
\label{fig:surveyarea}
\end{center}
\end{figure}

The distribution of the rms noise across the  mosaic is shown in Figure\,\ref{fig:rms_distribution}. We reach a point source sensitivity of $645\,\rm n Jy\,beam^{-1}$ at the pointing centers, with noise variations among these centers  less than 10\%. As also observed in the cumulative distribution of area {\it vs} rms noise  level (Figure\,\ref{fig:surveyarea}), the sensitivity remains nearly homogeneous within the central $\approx \, 120\,\rm arcmin^2$ region. Our mosaic extends beyond the area covered by the {\it HST} imaging of $170\,\rm arcmin^2$, albeit the sensitivity in such outer regions ranges from  $\rm 1 - 6 \,\mu Jy\,beam^{-1}$.

Finally, we produced a non-primary-beam corrected (“flat noise”) mosaic by reverting the weights (on a pixel-by-pixel basis) used to generate the primary-beam-corrected mosaic with Equation\,\ref{eq:mosaic}. This  mosaic facilitates the source extraction procedure and Monte Carlo simulations (see Section\,\ref{sec:cataloging} and \ref{sec:corrections}) as it prevents the presence of noisy edges. Likewise, it allows us to inspect the pixel brightness distribution in the mosaic without being affected by the primary beam attenuation. As observed in the left panel of Figure\,\ref{fig:histo_pixel_dist}, the noise amplitude distribution is fairly Gaussian with a clear excess of pixels with flux density above five times  the  rms noise  $\sigma_{\rm n}=\rm\, 671\, nJy\,beam^{-1}$.

In the following, we refer to the mosaic with a beam FWHM $0\farcs22$ as our high-resolution mosaic. This mosaic is fundamental to derive the structural measurements of high-redshift radio sources in the GOODS-N field.

\subsubsection{Low-Resolution Mosaic}
We produced a low-resolution,  ($u, v$)-tapered mosaic with a $1\farcs0$ synthesized beam following the same approach as in Section\,\ref{subsubsec:high-res_mosaic}, i.e., we imaged each pointing individually with {\tt tclean} using a  pixel scale of $0.25\,\rm arcsec/pixel$, applied primary beam corrections with {\tt widebandpbcor}, and combined the deconvolved images with {\tt linearmosaic} to get the low-resolution mosaic covering the same 297$\rm \,arcmin^2$ region as in the high-resolution one. In this case, after extensive testing, we adopt {\tt robust=2.0} to minimize the noise level. The point source sensitivity at the pointing centers of the $1\farcs0$  tapered mosaic is $\approx 920\,\rm n Jy\,beam^{-1}$ (right panel of Figure\,\ref{fig:rms_distribution}) and, similar to the high-resolution mosaic, the rms noise across the low-resolution mosaic is nearly homogeneous within the central 120$\rm \,arcmin^2$ (see Figure\,\ref{fig:rms_distribution} and \ref{fig:surveyarea}). 

\begin{figure*}
\begin{center}
\includegraphics[width=1.0\textwidth]{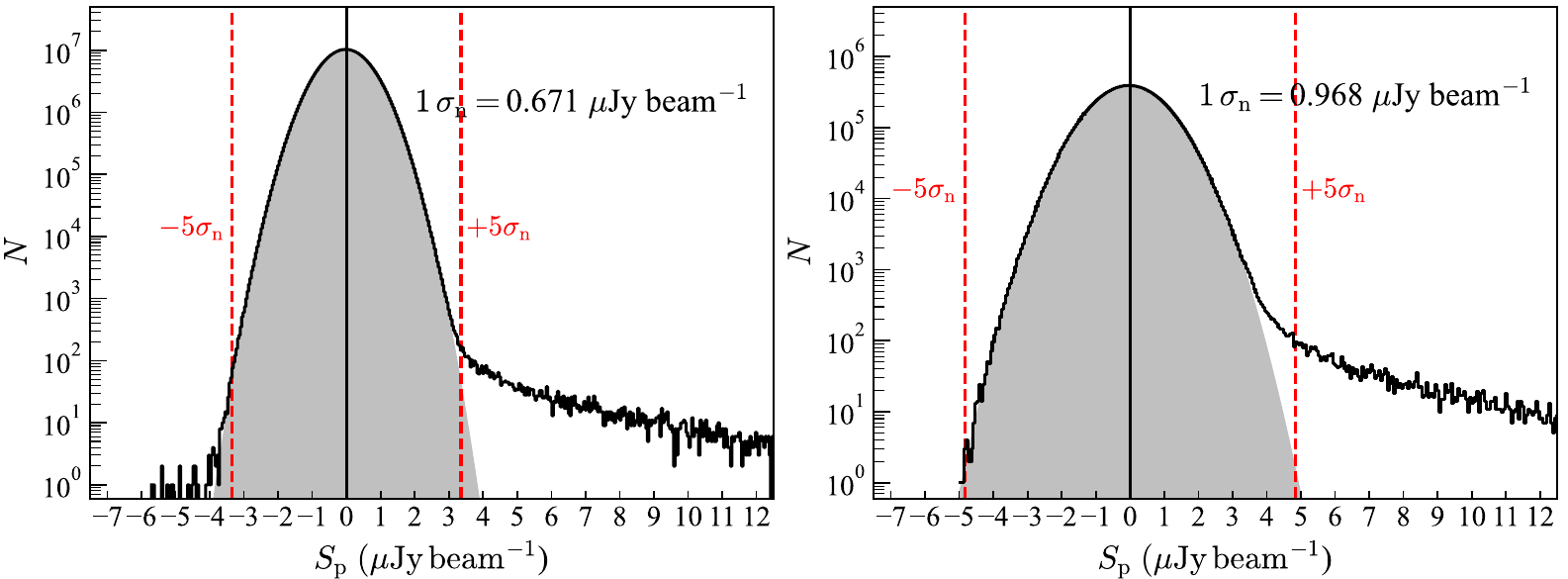}
\caption{Pixel brightness distribution of the 10\,GHz mosaic of the GOODS-N field at  $0\farcs22$ ({\it left}) and $1\farcs0$ resolution  ({\it right}). The pixel values shown here are  uncorrected for primary-beam attenuation.  The gray regions are Gaussian fits to the observed distributions. The rms noise  values are shown in the upper-right corners, while the $+/-5\sigma_{\rm n}$ levels are illustrated by the dashed vertical lines. The noise of both versions of the 10\,GHz mosaic of the GOODS-N field follows a Gaussian probability distribution, indicating that the mosaics are not limited by dynamic range.}
\label{fig:histo_pixel_dist}
\end{center}
\end{figure*}

\begin{figure*}
\centering
\includegraphics[width=.92\textwidth]{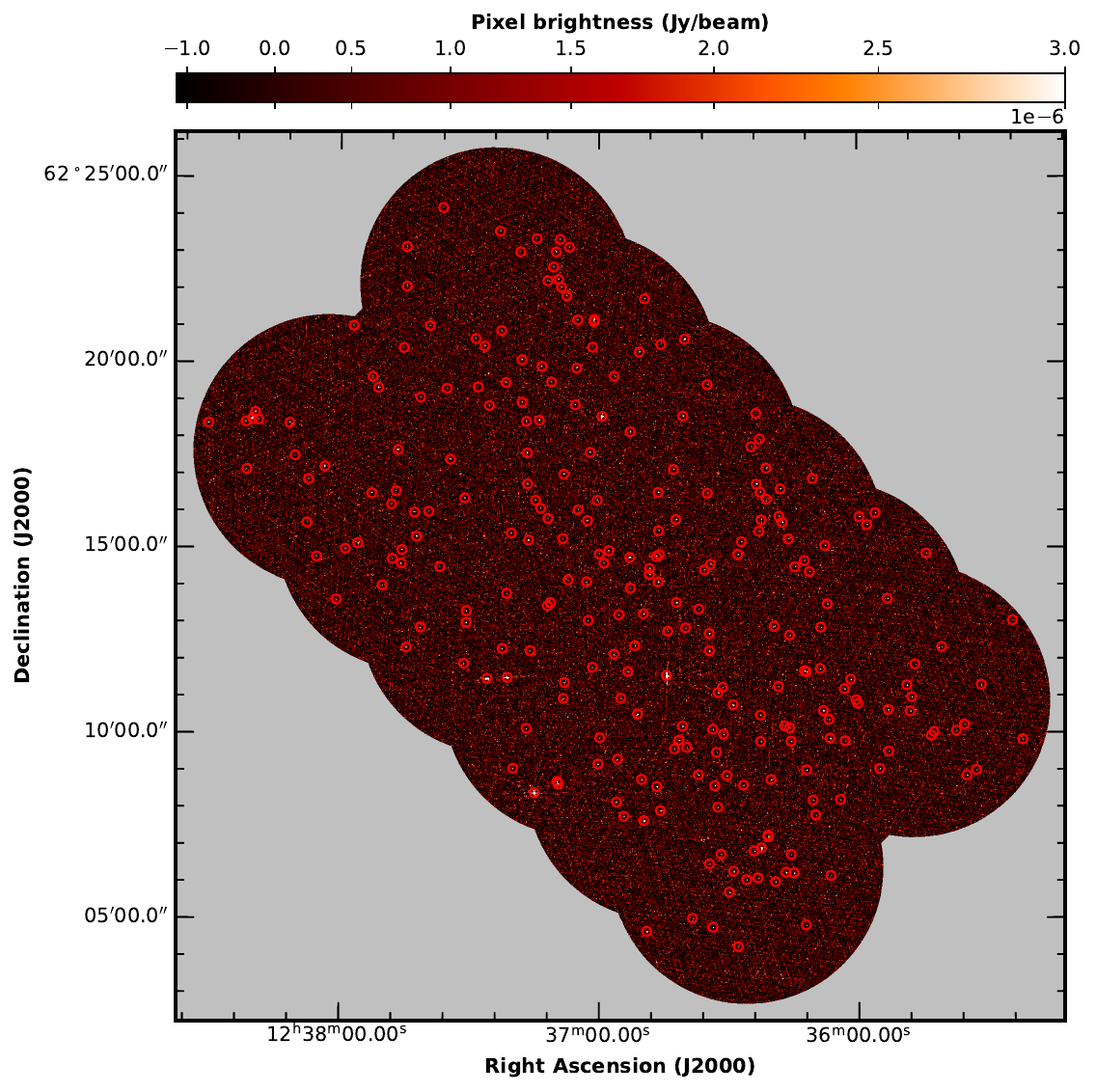}
\caption{The 10\,GHz mosaic of the GOODS-N field covering 297\,arcmin$^2$. Here, we show the non-primary-beam-corrected version of the mosaic at $1\farcs0$ resolution with a median rms noise   of 0.968\,$\rm \mu Jy\,beam^{-1}$. The red circles highlight the position of the 256 radio sources in our master 10\,GHz catalog of the GOODS-N field.}  
\label{fig:fullmap}
\end{figure*}

We also obtained a non-primary beam corrected (``flat noise'') version of this low-resolution mosaic (see Figure\,\ref{fig:fullmap}). Its pixel brightness distribution  is accurately described by a Gaussian function with $\sigma_{\rm n}=0.968\,\rm \mu Jy\, beam^{-1}$ (see right panel of Figure\,\ref{fig:histo_pixel_dist}). Besides, the distribution exhibits a clear excess of pixels with flux density values above five (and even three) times the noise amplitude, as well as the absence of pixels with negative values beyond the expected Gaussian distribution. 

While the rms noise of the low-resolution mosaic is 44\% higher than the high-resolution one, the corresponding brightness temperature rms of the low-resolution mosaic is 19 times lower than its high-resolution counterpart. As a result, the $1\farcs0$  tapered mosaic allows us to better probe the faint and extended radio emission of high-redshift radio sources. Moreover, as detailed in Section\,\ref{sec:cataloging}, the low-resolution map leads to a significantly lower fraction of spurious sources, rendering this  $1\farcs0$  tapered mosaic the preferred one to perform our blind source extraction and obtain the master radio source catalog.

\subsubsection{A Note About Joint Deconvolution and the A-Term Correction}

Given the large data volume of the survey, totaling $\approx$\,40\,TB of calibrated MSs, joint deconvolution was unfeasible even with  parallel processing recently implemented in {\tt mpicasa} and the computing resources available to us at NRAO. Likewise, we attempted correcting for the A-term to take into account the baseline, time, and frequency dependence of the aperture illumination pattern (AIP) of the antennas \citep{bhatnagar13}. This can be done with the gridder ``awproject'' in {\tt tclean} that applies the  A- and W-Projection algorithms. Nevertheless, the computing cost of the AW-Projection is significantly larger than standard imaging --even with parallelization.  Hence, joint deconvolution of our entire data set with the AW-projection was nonviable.  

To verify that the adopted approach to image our data set (i.e., imaging each pointing individually and combining them with {\tt linearmosaic}) leads to a mosaic that is consistent with the ones produced via joint deconvolution and the AW projection, we perform the following tests. We  downscale our data set and image only four out of the 17 pointings in our survey via joint deconvolution with the gridder ``mosaic'' and ``awprojection''.  
We find no significant difference between 
the pixel brightness distribution, rms noise, number of detected sources, fraction of spurious sources, and presence of imaging artifacts between the maps  produced via joint deconvolution with  gridder ``mosaic''$/$``awprojection'' and the map obtained with our adopted approach. Moreover, we find that the  structural parameters of detected sources (integrated flux density and major FWHM) in the three maps differ, in general, by $\lesssim 10\%$. 

{ Considering that the  next generation VLA  will be regularly producing radio surveys with ten times better angular resolutions and sensitivity than the 10\,GHz survey of GOODS-N \citep{murphy22},   it is worth stressing that massive computing resources will be needed to process such amount of data. Ongoing efforts to analyze and optimize the  size-of-computing  for  ngVLA synthesis imaging are being taken, concluding that parallelization and implementations based on Graphics Processing Units (GPUs) and Field Programmable Gate Arrays (FPGAs)  have the potential to reduce the computing costs of the next generation VLA\footnote{\url{https://library.nrao.edu/public/memos/ngvla/NGVLAC_04.pdf}}. For example, an experimental project led by the NRAO has shown that a nationwide grid of computers with GPUs can reduce the imaging running time by two orders of magnitude (S. Bhatnagar priv. comm.).}

% =================================================== 

\section{Catalog} 
\label{sec:cataloging}

We adopt the  low-resolution mosaic (with {\tt robust=2} and {\it uv}-tapered to a $1\farcs0$ resolution) to derive our master radio source catalog based on the following considerations.   

First,  after extensive tests, we find that a {\it uv}-tapered  map with a $1\farcs0$ resolution leads to the lowest fraction of spurious sources compared to (Section\,\ref{subsec:falsedetections})  any other map with higher resolution. { Note that  the pixel brightness distribution of the high-resolution mosaic (left panel of Figure\,\ref{fig:histo_pixel_dist}) shows a tail of negative
sources that deviates from the Gaussian model. These spurious sources are randomly distributed across the high-resolution mosaic and disappear in the low-resolution mosaic (right panel of Figure\,\ref{fig:histo_pixel_dist}).} Moreover, since using {\tt robust=2} leads to the best sensitivity, the low-resolution map is the best alternative to increase the number of detections while minimizing the presence of spurious sources. A detailed description of the tests performed to { unveil the dependence between the number of spurious sources and angular resolution of the map} will be part of an upcoming technical report/manuscript.

Second, a master catalog from the low-resolution mosaic  simplifies the implementation of Monte Carlo simulations to derive completeness and flux density boosting corrections (Section\,\ref{subsec:completeness} and \ref{subsec:fluxboosting}). Specifically,  mock radio sources ``observed'' at $1\farcs0$ resolution can be generated with a single 2D Gaussian model, instead of more complex models needed to reproduce the compounded radio structures that are revealed in the high-resolution mosaic with a  $0\farcs22$  resolution (see Figure\,\ref{fig:examples_cutout}). 

\subsection{Source Extraction}

\begin{figure}
\begin{center}
\includegraphics[width=1.0\columnwidth]{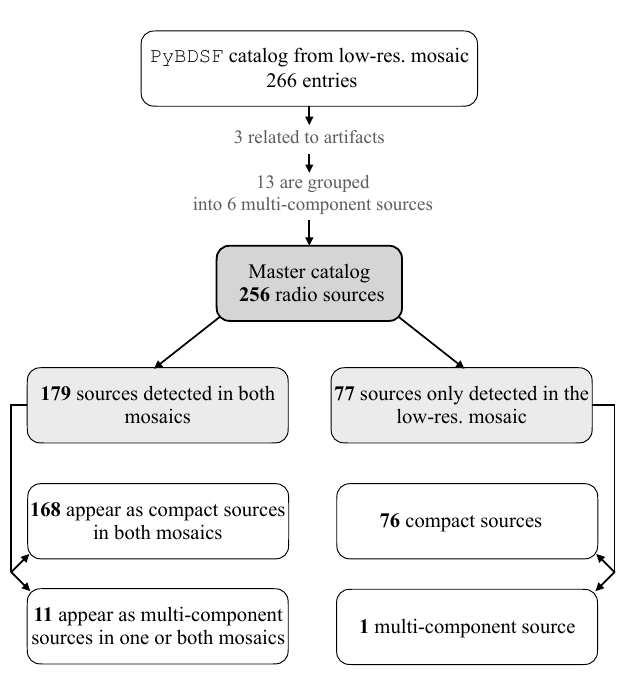}
\caption{A flowchart that outlines the process to obtain our master catalog from the low-resolution mosaic. The flowchart also illustrates the fraction of sources that have counterparts in the high-resolution mosaic and the fraction of multi-component sources in our master catalog.}
\label{fig:flowchart}
\end{center}
\end{figure}

 We use the Python Blob Detector and Source Finder \citep[{\tt PyBDSF};][]{Mohan15} to obtain  radio source catalogs. 
  The source extraction is performed in the ``flat noise'' mosaic to mitigate the effect of the noise edges on the calculation of the rms noise map, which is derived using the {\tt PyBDSF}´s suggested values for the box size and step size. We adopt a threshold to identify the islands of contiguous emission ({\tt thresh\_isl}) of 3$\sigma$, where $\sigma$ is the local rms noise.  The source detection threshold ({\tt thresh\_pix})  is set to  5$\sigma$. After visually inspecting the resulting  266 catalog entries, we find and remove three entries linked to artifacts around a bright radio source with $S_{\rm P}\approx \rm 410\,\mu Jy\,\rm beam^{-1}$. Also, we group 13 catalog entries into 6 multi-component, extended, and complex sources (see Section\,\ref{subsub:multicomponent-cat}). Our  master catalog, therefore, comprises  256 radio sources detected in the low-resolution mosaic (see Figure\,\ref{fig:flowchart}).

In addition, to provide more robust information on the major and minor FWHM of these radio sources, we run {\tt PyBDSF}  on the ``flat noise'' high-resolution mosaic using the detection thresholds and rms map derivation procedure adopted for our master catalog. By matching the catalogs from high- and low-resolution mosaics (using a $1\farcs0$ radius) and visually inspecting the radio sources, we find the following (see Figure\,\ref{fig:flowchart}). 

179 sources from our master catalog are detected with a peak signal-to-noise ratio SNR$\geq$5 in both the low- and high-resolution mosaic, of which 168 appear as compact sources in both mosaics, and 11 appear as multi-component radio sources in one or both of the mosaics.

\begin{figure}
\begin{center}
\includegraphics[width=.95\columnwidth]{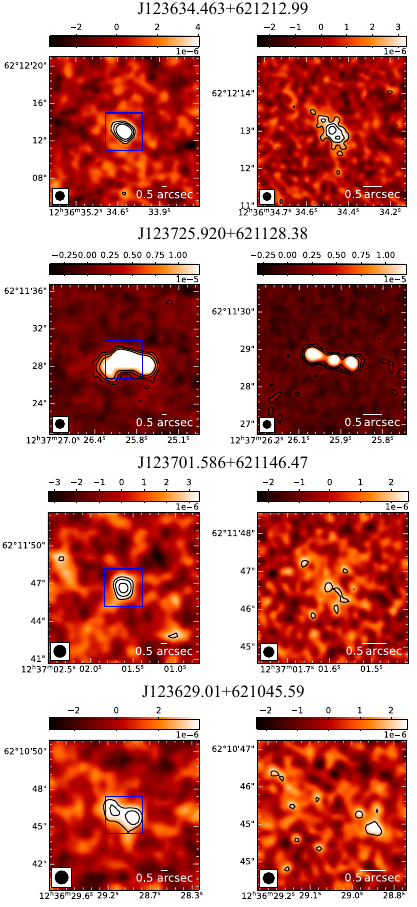}
\caption{Stamps from the 10\,GHz mosaic of the GOODS-N field at $1\farcs0$ ({\it left panels}) and $0\farcs22$\,arcsec resolution  ({\it right panels}).  The blue rectangle in the low-resolution stamp  highlights the zoomed-in region displayed in the stamp at $0\farcs22$ resolution. These  examples illustrate the four  types of radio sources in our master catalog from the low-resolution mosaic (see Figure\,\ref{fig:flowchart}): {\it (first/top row)} 168 compact radio sources with a single counterpart detected in the high-resolution mosaic, {\it (second row)} 11  sources that appear as multi-component in  the low- and/or high-resolution mosaic, {\it (third row)} 76 compact radio sources without a counterpart detected  in the high-resolution mosaic, {\it (fourth/bottom row)} 1 multi-component source without a counterpart detected  in the high-resolution mosaic. Contour levels are at 3, 5, and 8 times the rms noise. }
\label{fig:examples_cutout}
\end{center}
\end{figure}

77 sources from our master catalog are only detected in the low-resolution mosaic, of which 76  are compact sources, and one is a multi-component source.    In Figure\,\ref{fig:examples_cutout}, we present examples of the four types of radio sources in our master catalog listed above.

\begin{figure}
\begin{center}
\includegraphics[width=1.00\columnwidth]{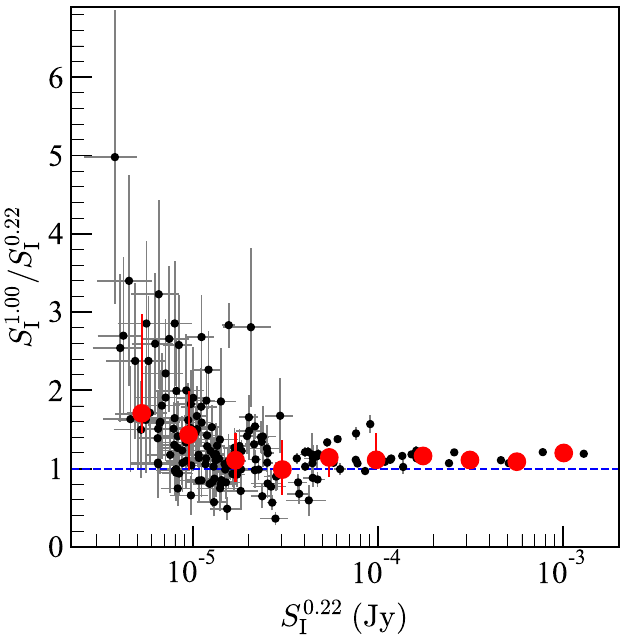}
\caption{Comparison between the integrated flux densities of 168  radio sources  detected in the high- and low-resolution mosaics of the  GOODS-N field. The blue dashed line is set at 1. At $S_{\rm I}^{0.22} > \rm 10\,\mu Jy$, the flux densities at low and high resolution typically are equal within $\approx 25\%$.
The red points show the median $S_{\rm I}^{1.00}/S_{\rm I}^{0.22}$ ratios  within a $\approx$0.25\,dex $S_{\rm I}^{0.22}$ bin. The error bars denote the 16th and 84th percentile values.}   
\label{fig:flux_comparison}
\end{center}
\end{figure}

{ Lastly, in Figure\,\ref{fig:flux_comparison}} we focus on the 168 compact radio sources detected in the low- and high-resolution mosaics and compare their integrated flux densities measured at both angular resolutions. In general, the flux densities derived at different resolutions are similar. At the faint end ($S_{\rm I}\lesssim \rm 10\,\mu Jy$), however, the low-resolution imaging allows us to retrieve a factor 1.5 more emission --on average-- than that observed in the high-resolution mosaic.

\subsection{Measuring Flux Densities of Multi-Component Radio Sources}
\label{subsub:multicomponent-cat}

To improve the flux density measurements of the  multi-component radio sources in our mosaics, we run {\tt PyBDSF} in the interactive mode and follow the software´s recommendations to fit extended and complex radio sources.  To improve the island determination,  we set {\tt rms\_map=False} and {\tt mean\_map=`const'} to use a constant mean and rms value across the fits file cutouts containing the multi-component radio sources.   We also set {\tt flag\_maxsize\_bm=50}  to fit larger Gaussian components when necessary, and {\tt atrous\_do=True} to fit Gaussians to the residual image and model any extended emission missed in the standard fitting. Finally, we adjust the  {\tt threshpix} parameter to improve the fitting and, if possible, to force {\tt PyBDSF} to associate multiple Gaussian components into a single, extended radio source.

\subsection{Radio Size Estimates}
\label{subsec:radiosizeestimates}
A total of 168 compact sources from our master catalog have a single counterpart in the high-resolution mosaic. Radio size estimates  for these 168 sources are derived from the high-resolution mosaic. In Table\,\ref{table:radiosourcecat}, we report the deconvolved FWHM ($\theta$)  along the major and minor axis  of the radio sources which, in the case of a circular beam, are given by:

\begin{equation}
\theta =\left( \phi^2- \theta_{1/2}^2 \right)^{1/2},
\end{equation}
where $\phi$ is the FWHM of the fitted major or minor axis of the source and $\theta_{1/2}$ is the FWHM of the synthesized beam. The uncertainties on the deconvolved FWHM that {\tt PyBDSF} reports are the same as the uncertainties on the FWHM values prior deconvolution. That seriously underestimates $\sigma_\theta$ for marginally resolved sources, so we estimate $\sigma_\theta$ using Equation (3) from \citet{murphy17} instead:

\begin{equation}
\left( \frac{\sigma_\theta}{\sigma_\phi}\right) =\left[1 - \left( \frac{\theta_{1/2}}{\phi}\right) ^2\right]^{-1/2}, 
\end{equation}
where $\sigma_\phi$ is the uncertainty on the fitted FWHM prior to deconvolution. In some cases, {\tt PyBDSF} reports unrealistic $\phi$ values (i.e., fitted FWHM equal to or smaller than the synthesized beam). The corresponding deconvolved FWHM are, therefore,  reported as  0 in Tables\,\ref{table:radiosourcecat} and \ref{table:multi-component_cat_0d2}, while the associated uncertainty corresponds to the  error on the fitted FWHM ($\sigma_\phi$).

Following \citet{murphy17}, we deem  sources meeting the criterion $\phi_{\rm M}-\theta_{1/2}\geq 2 \sigma_{\phi_{\rm M}}$ as confidently  resolved along their major (M) axis. Out of the 168 sources in our master catalog with a single counterpart in the high-resolution mosaic, 92 are confidently resolved.

In the case of sources that are not confidently resolved along their major (nor minor) axis, the peak brightness value approaches that of the integrated flux density. Therefore, we estimate  the geometric mean of the peak brightness and integrated flux densities   and adopt the resulting value as the best estimate for the integrated flux density ($S_{*}$; reported in Tables\,\ref{table:radiosourcecat} and \ref{table:multi-component_cat_0d2} as well). For sources whose major axes are resolved, the best estimate for the source´s integrated flux density  is simply that reported by {\tt PyBDSF}.

\subsection{Astrometric Accuracy}
\begin{figure}
\begin{center}
\includegraphics[width=1.0\columnwidth]{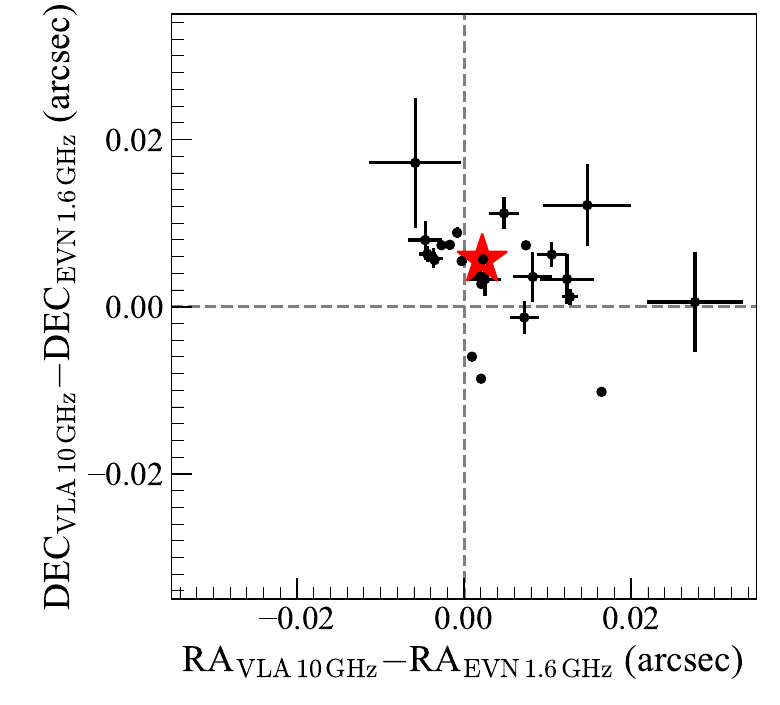}
\caption{Comparison between the central positions of 26 radio sources in our high-resolution VLA 10\,GHz mosaic and their counterparts in the EVN 1.6\,GHz data from \citet{radcliffe18}. The red star shows the median value from the offset distribution that is $\lesssim 5$\,milliarcsec. }
\label{fig:astrometry}
\end{center}
\end{figure}

We compare the positions of radio sources detected  with the European VLBI Network (EVN) at 1.6\,GHz \citep{radcliffe18} and their counterparts in our high-resolution VLA 10\,GHz mosaic of the GOODS-N field. Out of the 31 VLBI-detected sources, 26 are detected above peak $\rm SNR\gtrsim 10$ in our mosaic. The remaining five VLBI-detected sources either lie outside the footprint (four sources) or at the edge (one source) of the 10\,GHz mosaic --where the sensitivity drops by a factor of ten with respect to the central region. We find a median offset between the EVN 1.6\,GHz and VLA 10\,GHz positions of  only 2.3  milliarcsec in RA and 5.6 milliarcsec in DEC (see Figure\,\ref{fig:astrometry}). Using the positions from the low-resolution VLA 10\,GHz mosaic leads to median offsets of 29.7  milliarcsec in RA and 5.7 milliarcsec in DEC. The aforementioned mean positional offsets are $\lesssim 8$ times smaller than the pixel scale of the low- and high-resolution mosaics. We did not correct the 
catalogs' entries for the respective mean positional offsets derived here.

\subsection{Summary of Released Catalogs}
Our master source catalog comprises 256 radio sources detected in our $1\farcs0$ resolution mosaic of GOODS-N, out of which 12 are multi-component. The  flux densities of the 256 sources  (at both angular resolutions when available) are reported in Table\,\ref{table:radiosourcecat}.  Size estimates obtained from the $0\farcs22$ resolution mosaic are presented as well. In Tables\,\ref{table:multi-component_cat_1d0} and \ref{table:multi-component_cat_0d2},  we  report the properties  of the individual components/sources in the low- and high-resolution mosaics, respectively, that are grouped into the 12 multi-component  sources in our master catalog.

Henceforth, all the analyses to characterize our master catalog (Section\,\ref{sec:corrections}) and derive the radio source counts (Section\,\ref{sec:counts}) are based on the 1$\farcs0$ resolution mosaic of GOODS-N---from which our master radio source catalog is extracted.  \\

\begin{figure*}
\begin{center}
\includegraphics[width=1.0\textwidth]{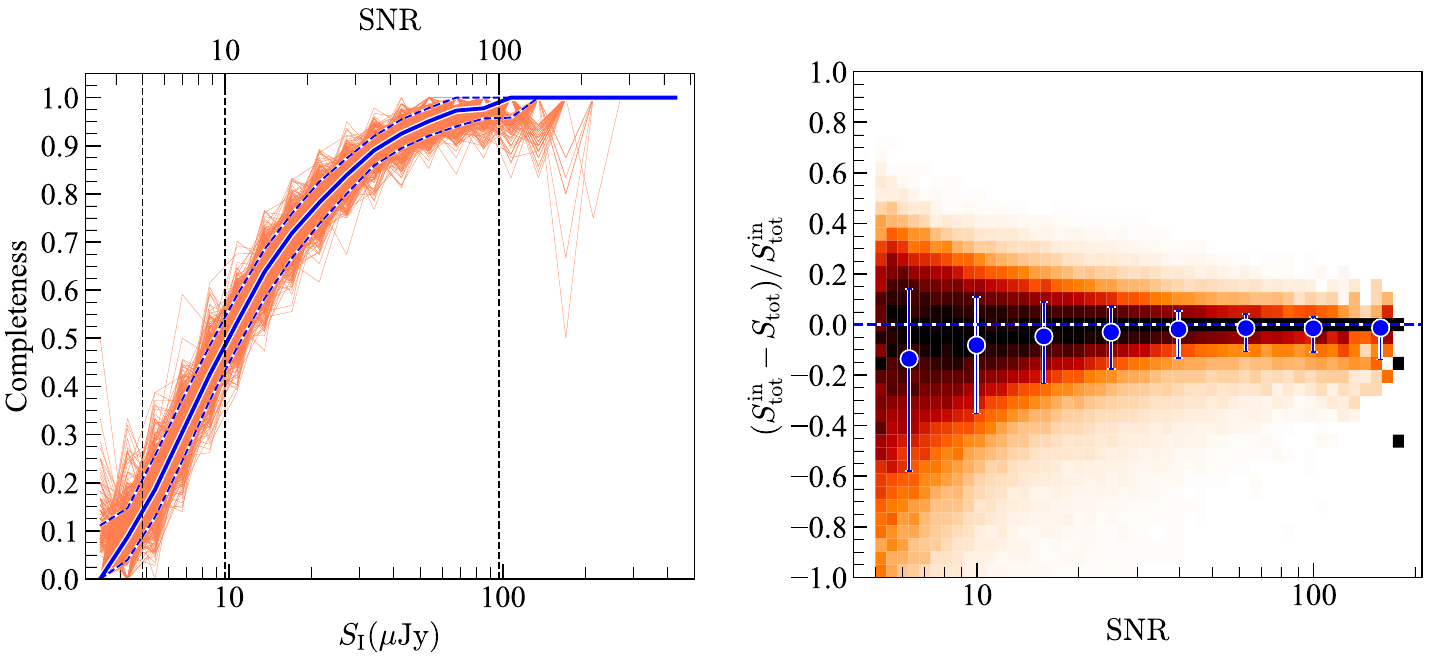}
\caption{{\it Left:} Completeness of the 10\,GHz source catalog of the GOODS-N field as a function of integrated flux density and SNR (for unresolved sources). The thin red lines show the completeness curve for each of the 500 mock catalogs in our Monte Carlo simulations. The solid and dashed blue lines show the 50th and 16th/84th percentiles, respectively.  {\it Right:} Flux boosting as a function of SNR of mock sources in our Monte Carlo simulations. A 2D histogram is shown in the background. The darker colors indicate a higher density of sources per flux density bin. The blue markers show the  50th percentile of the input-to-output flux density ratios per 0.2\,dex bin, while the error bars represent the 16th and 84th  percentiles. }
\label{fig:mc_simulaitons}
\end{center}
\end{figure*}

 \section{Radio source counts corrections} 
\label{sec:corrections}
We assess the reliability of the master 10\,GHz radio source catalog  of GOODS-N by deriving corrections factors  to account for the completeness, flux boosting, false detections, and resolution bias.

 \subsection{Completeness} 
 \label{subsec:completeness}
 To determine the number of sources that exist in a given region of the sky (above a detection limit) but are missed in our mosaic/catalog due to the adopted observational and detection procedure, we perform extensive Monte Carlo simulations as follows. 

 \begin{enumerate}
     \item  We  infer the probability distribution function (PDF) of the peak brightness ($S_{\rm P}$) and deconvolved major/minor FWHM ($\theta_{\rm M,m}$) of the sources in our catalog by fitting  an exponential  and a half-norm function ($f(x)=\sqrt{2/\pi} \exp(-x^2/2)$; for $x\geq 0$), respectively. 
     
     \item We use the inferred PDFs of the $S_{\rm P}$ and $\theta_{\rm M,m}$ to generate a mock sample of radio sources.    We verify that the distribution of the integrated flux density from the mock catalog matches, in general, that of the observed catalog. 

     \item 1000 mock radio sources are injected in the mosaic at random positions and position angles, under the condition that mock sources are located $\approx 9\,\rm arcsec$ away from  real  or other mock sources. These mock sources follow a $S_{\rm P}$ distribution that includes values as low as $3\mu \rm Jy\,beam^{-1}$, i.e., $\approx 3\times$ the average rms noise of our mosaic. By injecting these faint sources we consider the effect of the noise in boosting/decreasing the flux density of  sources  with $\rm SNR \approx 5$.

    \item We take into account  the rms noise variations in our mosaic, { mainly arising from the primary beam attenuation of the 17 pointings}, to derive our mock catalogs and completeness corrections as follows. Let us first gauge an illustrative case of a compact source with $S_{\rm I}\approx 50\,\rm \mu Jy$ that lies at the edge of our mapped region. While compact sources with such $S_{\rm I}$ values  are robustly detected with a $\rm SNR\approx  50$ in the central region of our mosaic, a source with $S_{\rm I}\approx \rm 50\mu Jy$  is detected with  $\rm SNR\approx 5$ at the outskirts of the map where the primary beam response drops to 10\%, which hinders the completeness of our catalog for such a hypothetical flux density value. Therefore, to  { consider the effect of the primary beam response in our completeness correction},  the input flux density of the mock sources is reduced depending on their position in the mock mosaic.

    \item We repeat  steps two to four and generate 500 mock mosaics, translating into a half million sources in our Monte Carlo simulations. Then, we obtain the corresponding 500 mock catalogs with {\tt PyBDSF} by applying the same detection parameter criteria used  to obtain our 10\,GHz catalog of the GOODS-N field.  

    \item We derive the completeness of our catalog by comparing the number of detected  sources with the number of injected  sources, per integrated flux density bin,  for all the mock mosaics/catalogs. The resulting 500 completeness curves, { already corrected by the primary beam attenuation across the mosaic},   are shown in the left panel of Figure\,\ref{fig:mc_simulaitons}. We adopt the median trend and the 16th/84th percentiles as our best completeness values ($C_{\rm comp}$) and associated errors, which are also reported in Table 1.  

 \end{enumerate}

\begin{deluxetable}{c  l l r r }
\tabletypesize{\footnotesize}
\tablecolumns{3}
\tablecaption{Completeness correction factors ($C_{\rm comp}$) for the GOODS-N 10\,GHz catalog as a function of integrated flux density.   \label{table:completeness}}
\tablehead{	\colhead{Integrated flux density} & \colhead{$C_{\rm comp}$}  & \colhead{Error}   \\
\colhead{($\mu \rm Jy$)} & \colhead{ }  & \colhead{ } }  
\startdata
$\leq$3.4	&	0.00	&	0.00	 \\
4.3	&	0.09	&	0.05	 \\
5.4	&	0.18	&	0.06	 \\
6.8	&	0.30	&	0.06	 \\
8.5	&	0.42	&	0.06	 \\
10.7	&	0.53	&	0.05	 \\
13.5	&	0.64	&	0.05	 \\
17.0	&	0.72	&	0.05	 \\
21.4	&	0.78	&	0.04	 \\
26.9	&	0.84	&	0.04	 \\
33.9	&	0.89	&	0.03	 \\
42.7	&	0.92	&	0.03	 \\
53.7	&	0.95	&	0.03	 \\
67.6	&	0.97	&	0.03	 \\
85.1	&	0.98	&	0.02	 \\
107.1	&	1.00	&	0.00	 \\
134.9	&	1.00	&	0.00	 \\
169.8	&	1.00	&	0.00	 \\
213.8	&	1.00	&	0.00	 \\
269.1	&	1.00	&	0.00	 \\
338.8	&	1.00	&	0.00	 \\
426.6	&	1.00	&	0.00	 \\ \hline 
\enddata
\end{deluxetable}

 \subsection{Flux Boosting}
 \label{subsec:fluxboosting}

 To characterize the effects of flux boosting \citep[e.g.,][]{coppin05, casey14} in our sensitivity-limited mosaic, we contrast the input and output flux density ($S_{\rm I}^{\rm in}$ and $S_{
 \rm I}$, respectively) of the mock sources in our Monte Carlo simulations.  We find that flux densities are boosted by $\approx 14\%$ in the lowest SNR bin centered at $S_{\rm I}\approx 6\mu \rm  Jy$ (Figure\,\ref{fig:mc_simulaitons}). At SNR larger than  15, flux densities are boosted by less than 5\%. Consequently, we do not apply this correction to the flux densities in the catalog.

 \subsection{False Detections}
  \label{subsec:falsedetections}

 We determine the fraction of spurious sources in our catalog by performing the source extraction with {\tt PyBDSF} on the inverted (i.e., multiplied by -1) mosaic. To this end, we use the same detection parameters adopted to obtain the 10\,GHz catalog of GOODS-N. Two spurious sources  are detected with $\rm SNR\approx 5$, leading to a total fraction of spurious sources in our catalog of only 0.75\%. { This implies a notably high   ``fidelity'' parameter of 0.99, defined as $1-N_{\rm neg}/N_{\rm pos}$ with $N_{\rm neg}$ and $ N_{\rm pos}$ the number of negative and positive detections, respectively \citep{decarli20}. } 
 
 Comparing the number of spurious sources with the number of sources detected in our mosaic per SNR bin, we find a fraction of spurious sources of 4\% within $5.0 \leq \rm SNR < 5.5$, and  0\% for SNR larger than 5.5. These sources are detected at the outskirts of the map, where the primary beam response is 0.1414 and 0.4359, and have  a total (primary beam-attenuated) integrated flux density  of $6.59\pm 2.18 
\rm \mu Jy$ and $9.69\pm 2.90 \rm \mu Jy$, respectively. Considering the primary-beam corrected flux densities, the fraction of spurious sources ($f_{ss} $) is zero in all  $S_{\rm I}$  bins except that spanning from  15 to $50\rm \mu Jy$ where $f_{ss} =0.02$ (see Table\,\ref{table:source_counts}). 

 \subsection{Resolution Bias}
  \label{subsec:resolutionbias}

Detecting sources in our SNR thresholded mosaic relies on the peak brightness. An unresolved source in our mosaic with $S_{\rm I}\approx\rm  10\mu Jy$, for example, has a greater probability of being detected than an extended source with the same $S_{\rm I}$ value but lower peak brightness. This effect  will hinder the number of detections, particularly close to our detection limit. As a result, the population of extended, low surface brightness sources might be underrepresented in our original catalog and mock mosaics.   We address this so-called ``resolution bias'' by following the analytic  methodology presented in \citet{vandervlugt21} that is summarized in the following lines. 

The resolution bias correction factor ($f_{\rm rb}$) is given by
    \begin{equation}
        f_{\rm rb}=\left[  1- h(>\theta_{\rm max}) \right]^{-1},
        \label{eq:resol_bias}
    \end{equation}
where $h(>\theta_{\rm max})$ is the fraction of sources expected to be larger than the  maximum angular size ($\theta_{\rm max}$) that our detection procedure is sensitive to. Such a fraction can be inferred with the relation \citep{windhorst90}: 
    \begin{equation}
        h(>\theta_{\rm max}) =  \exp\left[ -\ln(2) \left( \frac{\theta_{\rm max}}{\theta_{\rm med}} \right)^{0.62} \right]. 
    \end{equation}
$\theta_{\rm max}$  depends on the source´s integrated flux density and  is expressed as 
    \begin{equation}
        \theta_{\rm max} = \left[ \theta_{\rm 1/2}^2  \times (S_{\rm I}/5\sigma) \right]^{1/2},
    \end{equation}
with  $\theta_{\rm 1/2}$ the FWHM of our synthesized circular  beam. Finally, $\theta_{\rm med}$ is the median angular size of the radio source population. \citet{windhorst90} propose a flux-dependent size given by $\theta_{\rm med}= 2(S_{\rm 1.4\,GHz})^{0.3}$, where  $S_{\rm 1.4\,GHz}$ is the flux density in millijanskys that we infer by scaling our 10\,GHz measurements with a spectral index of $-0.8$. We also adopt a constant value of  $\theta_{\rm med}=0\farcs30$ that has been specifically derived for the $\mu \rm Jy$ radio source population \citep{cotton18, bondi18}. By evaluating Equation\,\ref{eq:resol_bias}  we derive $f_{\rm rb}$ for a flux-dependent and constant  $\theta_{\rm med}$. Here, we adopt the mean of these two values as our best estimate for the  resolution bias correction factor, which is $\approx1.40$ for integrated flux densities  $\approx 5-15\mu \rm Jy$ and less than 1.20 for  $S_{\rm I}\gtrsim 100\mu \rm Jy$.  A list with the $f_{\rm rb}$ values per $S_{\rm I}$ bin can be found in Table\,\ref{table:source_counts}.

 \section{Radio source counts} 
 \label{sec:counts}

To derive the observed Euclidean-normalized number counts, following \citet{matthews21},  we first estimate the quantity

\begin{equation}
    S_{\rm I}^2 n(S_{\rm I})=\left[ \frac{1}{\Omega \ln(\Delta)} \right] \sum^{n_{\rm bin}}_{i=1} S_{\rm I}^i,
    \label{eq:source-counts}
\end{equation}
where $n_{\rm bin}$ is the number of sources per $S_{\rm I}$ bin, $\Omega$ is the survey area of $297\,\rm arcmin^2$, and $\Delta$ the  logarithmic width of the bin of $\rm dex(0.5)$. We also derive the rms statistical uncertainty in $S_{\rm I}^2 n(S_{\rm I})$  using  

\begin{equation}
   \sigma_{\rm stat}=\left[ \frac{1}{\Omega \ln(\Delta)} \right] \left( \sum^{n_{\rm bin}}_{i=1} \left( S_{\rm I}^i \right)^2 \right)^{1/2},
    \label{eq:e-source-counts}
\end{equation}
{  which is valid for bins with $n_{\rm bin}\gg 1$. We note that the uncertainty for the brightest bin (with $n_{\rm bin}=4$) is potentially larger than the quoted/plotted value due to Poissonian fluctuations. } 

\begin{figure*}
\begin{center}
\includegraphics[width=.65\textwidth]{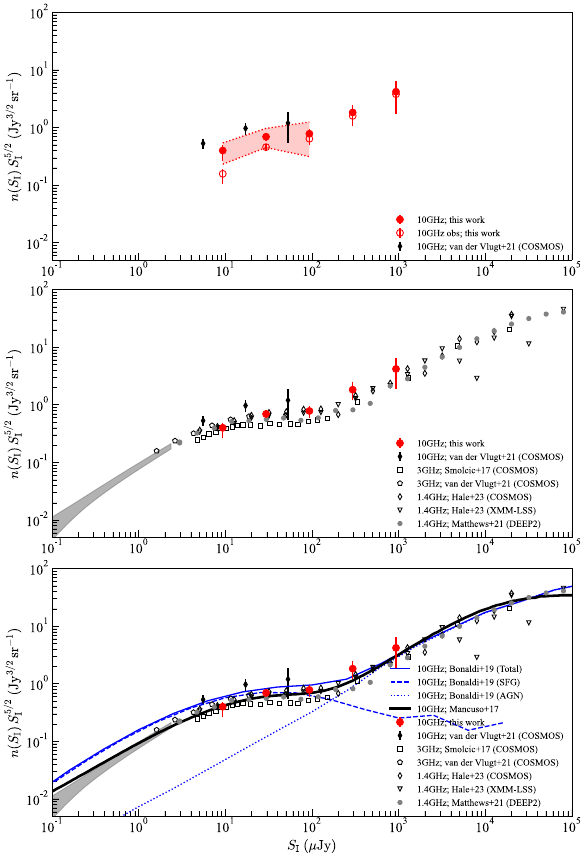}
\caption{{\it Top panel:} Euclidean-normalized differential radio source counts at 10\,GHz from the GOODS-N field (filled red circles), which have been corrected by completeness, fraction of spurious sources, and resolution bias. The uncorrected source counts are shown as open red circles. These are the first  radio source counts obtained at high frequencies ($\nu \gtrsim 5\,\rm GHz$) from an entire extragalactic deep field. For comparison, we also present the radio source counts at 10\,GHz from a single-pointing  VLA map in the COSMOS field \citep{vandervlugt21}. To illustrate the effect of cosmic variance on single-pointing  VLA imaging at 10\,GHz,  we present the maximum/minimum radio source counts (red dotted lines) obtained from six  nonoverlapping subfields each covering 32.16\,$\rm arcmin^2$ in our 10\,GHz mosaic of GOODS-N. 
{\it Middle panel:} A compilation of radio source counts measured at 1.4 and 3\,GHz \citep{smolcic17, matthews21,vandervlugt21,  hale23}, including the  statistical counts derived from a $P(D)$ analysis at 1.4\,GHz in the DEEP2 field \citep[shaded gray region;][]{matthews21}. The 1.4 and 3\,GHz radio source counts are shifted to the 10\,GHz observed frame by adopting a spectral index of $-0.7$, to facilitate the comparison between 10\,GHz radio source counts \citep[][and this work]{vandervlugt21} and the expected values from the more abundant 1.4 and 3\,GHz measurements. No error bars are shown here to aid the visual inspection of the plot. {\it Bottom panel:} Radio source counts measured at and/or shifted to 10\,GHz compared with the simulations and model from \citet{mancuso17}   and \citet{bonaldi19}, who made specific predictions for the radio sky population at the $\approx$10\,GHz observed frame. The dotted and  dashed blue lines show the radio source counts of the AGN and SFG populations, respectively, as derived by \citet{bonaldi19}.  }
\label{fig:sourcecounts}
\end{center}
\end{figure*}

Then, the observed Euclidean-normalized number counts are estimated by using $S_{\rm I}^2 n(S_{\rm I})$ multiplied by $S_{\rm I}^{1/2}$. These number counts must be corrected for completeness, fraction of spurious sources, and resolution bias. Thus, to derive the corrected Euclidean-normalized number counts we calculate 

\begin{equation}
    S_{\rm I}^2 n(S_{\rm I})=\left[ \frac{1}{\Omega \ln(\Delta)} \right] \sum^{n_{\rm bin}}_{i=1}   \frac{[1-f_{\rm ss} (S_{\rm I}^i)] \times  f_{\rm rb} (S_{\rm I}^i)}{C_{\rm comp}(S_{\rm I}^i)},
        \label{eq:cor-source-counts}
\end{equation}
where $C_{\rm comp}$, $f_{\rm ss}$, and  $f_{\rm rb}$  are the aforementioned completeness, spurious sources, and resolution bias correction factors, respectively. The observed and corrected  Euclidean-normalized 10\,GHz radio source counts of the GOODS-N field are shown in the top panel of Figure\,\ref{fig:sourcecounts}.  These cover the flux density range  $-5.28<\log(S_{\rm I}/\rm Jy)<-2.78$   in five bins of 0.5\,dex-width. In Table\,\ref{table:source_counts}, we also report the correction factors and radio source counts per flux density bin.

We compare the 10\,GHz source counts with those derived by \citet{vandervlugt21} using an ultra-deep, single-pointing VLA data in the COSMOS field.  Their X-band VLA image reaches an rms noise level of $0.41\, \rm \mu  Jy\,beam^{-1}$ at the pointing center and has an angular resolution of $2\farcs33\times 2\farcs01$. As observed in the  top panel of Figure\,\ref{fig:sourcecounts}, the 10\,GHz radio source counts derived by \citet{vandervlugt21} span over   $-5.57<\log(S_{\rm I}/\rm Jy)<-4.10$, allowing a more direct comparison between the 10\,GHz number counts derived in both studies across $-5.28<\log(S_{\rm I}/\rm Jy)<-4.10$. Within this flux density regime, the 10\,GHz number counts from \citet{vandervlugt21} are systematically higher by a factor $\approx 1.6$ than those reported here. { We discuss this discrepancy within the context of sample and cosmic variance in Section \ref{subsec:cosmic-variance}}. 
 
\subsection{Comparison with 1.4 and 3\,GHz Number Counts}
\label{subsec:comparisocounts}

We compare the 10\,GHz radio source counts in the GOODS-N field with the more abundant measurements obtained at 1.4 and 3\,GHz in the COSMOS, XMM Large Scale Structure (XMM-LSS), and DEEP2 fields \citep{smolcic17, matthews21,vandervlugt21,  hale23}.  To this end, the  1.4 and 3\,GHz radio source counts are converted to the 10\,GHz observed frame  assuming that the radio SED is described by a  power-law: $ S_{\rm \nu } / S_{\rm 10} = ( \nu \,[\rm GHz]/ 10 )^\alpha $, where $\alpha$ is the spectral index that here is fixed to $-0.7$. Then, the 10\,GHz radio source counts ($n_{\rm 10}$)  are estimated from the 1.4 and 3\,GHz values ($n_\nu$) using  $n_{\rm 10}= n_{\nu} ( 10/ \nu\, [\rm GHz])^{1.5\alpha}$. As observed in  the middle panel of Figure\,\ref{fig:sourcecounts}, the 10\,GHz radio source counts from GOODS-N follow, in general,  the trend depicted by the  10\,GHz number counts inferred from the lower frequency observations. Radio source counts rise from the sub-$\mu \rm Jy$ regime, they flatten out at flux densities  $ 10 \,{\rm \mu Jy} \lesssim S_{\rm I} \lesssim 100 \, \rm \mu Jy$, and then continue to rise towards the bright end. The scatter of the number counts from all these studies is, however, evident. This can be attributed to the different assumptions made to correct for the resolution bias and completeness, as well as field-to-field variations due to sample and cosmic variance  \citep[as also discussed by][]{smolcic17, vandervlugt21, hale23}. Furthermore, the small discrepancies between the different trends observed in the middle panel of Figure\,\ref{fig:sourcecounts} are also a consequence of the simplistic assumptions made here to scale the 1.4 and 3\,GHz radio source counts to the 10\,GHz observed frame, i.e., a single power-law radio SED with $\alpha=-0.7$. We note, however, that adopting flatter or steeper spectral indices  alleviates the discrepancies between the  scaled 1.4/3\,GHz and observed 10\,GHz radio source counts across different flux density regimes. For example, using  $\langle \alpha^{10\,\rm GHz}_{\rm 1.4\,GHz} \rangle=-0.61$---as reported for 10\,GHz radio sources in our pilot survey with and without a counterpart at 1.4\,GHz \citep{murphy17}---increases the normalization of the scaled 1.4\,GHz source counts by $\approx 0.12$\,dex. This leads to an excellent agreement between the robustly constrained 1.4\,GHz source counts in DEEP2 \citep{matthews21} and our 10\,GHz estimates in GOODS-N for flux densities $ 16 \, \rm \mu Jy \lesssim S_{\rm I} \lesssim 166 \, \rm \mu Jy  $. However,  a steeper spectral index of $\langle \alpha^{10\,\rm GHz}_{\rm 1.4\,GHz} \rangle \approx -0.8$ is needed to get a better agreement between   the scaled 1.4 and 10\,GHz number counts for flux densities $S_{\rm I} \lesssim 16 \, \rm \mu Jy$, which is compatible with the average spectral index steeper than $-0.75$ found for 6 and 8.5\,GHz-detected sources below 35$\mu$Jy \citep{fomalont02, thomson_19}.

\subsection{The Impact of Sample and/or Cosmic Variance on Radio Source Counts}
\label{subsec:cosmic-variance}
{ As mentioned before, sample and/or cosmic variance could be one of many factors driving the scatter of the number counts obtained from different radio surveys. To illustrate this, we focus on a comparison between  the single- and multiple-pointing 10\,GHz data in the COSMOS \citep{vandervlugt21} and GOODS-N field, respectively. Both data sets have been obtained with the X-band receivers of  the VLA, reached similar depths and comparable angular resolutions, leaving the survey area as the main variable in our comparison.

We start by splitting} the  10\,GHz mosaic of GOODS-N into six nonoverlapping regions each covering $32.16\, \rm arcmin^{2}$, matching the area of the  single-pointing 10\,GHz map of COSMOS from \citet{vandervlugt21}.  Radio source counts from these six subfields are then obtained and corrected following Equation\,\ref{eq:cor-source-counts}, as done for the radio source counts from the full 10\,GHz mosaic of GOODS-N. The maximum and minimum number counts from these subfields, shown in the top panel of Figure\,\ref{fig:sourcecounts}, suggest that sample variance induces an additional scatter of $\rm \approx 0.15-0.40\,dex$  in the radio source counts measured in areas as small as $32.16\, \rm arcmin^{2}$. We note that such a scatter should be considered as a lower limit to the true cosmic variance arising from the large-scale structure of the Universe because the cosmic variance in the volume probed by the GOODS-N field is significant \citep[$\approx 10-30\%$ at redshifts $0<z<4$,][]{somerville04, driver10}. The systematically higher number counts reported in \citet{vandervlugt21}, therefore,  could be a result of sample and/or cosmic variance due to the relatively small region covered by the single-pointing VLA data. Additionally, given the two times coarser angular resolution of the COSMOS  data, the resolution bias (see Section\,\ref{subsec:resolutionbias}) could be also driving the higher number counts; even though we adopt the same method used by \citet{vandervlugt21} to correct the number counts from the potentially missing population of extended, low surface brightness sources.

\subsection{Comparison with Numerical Simulations and Models}
\label{subsec:comparisonmodels}

A more direct comparison between the 10\,GHz radio source counts from GOODS-N can be made with the numerical simulations and models reported by \citet{mancuso17}  and \citet{bonaldi19}, who provide specific predictions for the radio continuum sky at $\approx$10\,GHz (bottom panel of Figure\,\ref{fig:sourcecounts}). First, \citet{mancuso17} employ redshift-dependent SFR functions that are mapped into bolometric AGN luminosity functions, using deterministic evolutionary tracks for the star formation and supermassive black hole accretion in an individual galaxy, to  derive differential number counts at 150\,MHz, 1.4\,GHz, and 10\,GHz. Second, \citet{bonaldi19} present the Tiered Radio Extragalactic Continuum Simulation (T-RECS), which links the radio source positions  to those of dark matter halos in the  P-Millennium simulation \citep{baugh19}. Covering a 25\,deg$^2$ field of view, T-RECS employs redshift-dependent AGN and SFR functions to produce a set of simulated catalogs covering the frequency range from 150 MHz to 20 GHz. The simulated 10\,GHz number counts presented here are obtained by interpolating the data points from the T-RECS catalogs at 920 and 1250\,MHz.

As observed in the bottom panel of Figure\,\ref{fig:sourcecounts}, our 10\,GHz number counts at flux densities  $\gtrsim 16\,\rm \mu Jy$ are consistent with \citet{mancuso17}'s  and \citet{bonaldi19}'s  predictions within $\approx 1.2\,\sigma$.
Yet,  the number counts in our faintest bin,   where the dominant radio source population are SFGs (see bottom panel of Figure\,\ref{fig:sourcecounts}),  are in  better agreement with the predictions from  \citet{mancuso17}.  The 10\,GHz source counts at $S_{\rm I}\approx 10\rm \, \mu Jy$ from \citet{bonaldi19}  are a factor  $\approx1.5$ higher than those reported here and the predictions from \citet{mancuso17}, 
which could be a result of the assumptions made  to simulate the SFG populations. While \citet{bonaldi19} adopts an evolving FIR-radio correlation and a modified Schechter parameterization (i.e., with two characteristic slopes) to model the SFR function, \citet{mancuso17} do not vary the FIR-radio ratio and employs a standard Schechter function.

\begin{deluxetable*}{c  c c c c c c c }
\tabletypesize{\footnotesize}
\tablecolumns{3}
\tablecaption{Euclidean-normalized differential 10\,GHz radio source counts in the GOODS-N field. \label{table:source_counts}}
\tablehead{	\colhead{$\Delta S_{\rm I}$} & \colhead{$S_{\rm I}$}  & \colhead{$N$}  & \colhead{$n(S_{\rm I}) S_{\rm I}^{5/2}$} & \colhead{$f_{\rm ss}$} & \colhead{$f_{\rm rb}$} & \colhead{$C_{\rm comp}$}  \\
\colhead{($\mu \rm Jy$)} & \colhead{($\mu \rm Jy$) }  & \colhead{ } &  \colhead{($\rm Jy^{3/2}\,sr^{-1}$) }  & & & }  
\startdata
5.24-16.56	&	9.31	&  128 & $0.40^{+0.09}_{-0.13}$ & 0.00 & $1.37^{+0.05}_{-0.03}$ & $0.57^{+0.12}_{-0.18}$	\\
16.56-52.38	&	29.45	&  92 &  $0.69^{+0.11}_{-0.10}$ & 0.02 & $1.30^{+0.02}_{-0.03}$  & $0.81^{+0.09}_{-0.08}$	\\
52.38-165.63	&	93.14 & 22 & $0.78\pm0.18$ & 0.00 &          $1.21^{+0.02}_{-0.03}$ & $0.98^{+0.02}_{-0.04}$	\\
165.63-523.77    &  294.5 & 10 & $1.83\pm0.62$ & 0.00 &          $1.15^{+0.01}_{-0.02}$ & 1.0	\\
523.77-1656.29 & 931.40	&  4 &   $4.20\pm2.30$ & 0.00 &          $1.11^{+0.01}_{-0.02}$  &  1.0	\\
\enddata
\tablecomments{$\Delta S_{\rm I}$ is the flux bin, centered at $S_{\rm I}$,  within which the radio source counts $n(S_{\rm I})$ are estimated using  $N$ number of sources per bin. The listed Euclidean-normalized differential 10\,GHz radio source counts have been corrected using the $f_{\rm ss}$, $f_{\rm rb}$, and $C_{\rm comp}$ factors that account for the  fraction of spurious sources, resolution bias, and completeness, respectively (see Equation\,\ref{eq:cor-source-counts}). The quoted errors in the corrected source counts are estimated by quadratically adding the fractional errors in  $f_{\rm fb}$ and $C_{\rm comp}$ and the fractional rms statistical uncertainty from Equation\,\ref{eq:e-source-counts}}.  
\end{deluxetable*}

 \section{Summary and conclusions} 
 \label{sec:conclusions}
We have presented the first radio continuum survey ever obtained in an entire extragalactic deep field at high frequencies: the 10\,GHz survey of GOODS-N. The main data products  derived from this VLA Large Program\footnote{The radio continuum mosaics and catalogs are available at \url{https://science.nrao.edu/science/surveys/vla-x-gn/home}}, as well as the results from the inferred 10\,GHz radio source counts, are summarized below.  

\begin{itemize}

  \item Two versions of the mosaic covering an area of $297\,\rm arcmin^{2}$ have been produced. One high-resolution mosaic with a synthesized circular beam with FWHM $0\farcs22$  and point-source sensitivity of $671\rm \,nJy\,beam^{-1}$, and a low-resolution, ({\it u,v})-tappered mosaic with an angular resolution of $1\farcs0$  and  $968\rm \,n Jy\,beam^{-1}$ depth.

   \item  We have adopted the low-resolution mosaic to obtain our master 10\,GHz catalog of the GOODS-N field, which comprises 256 radio sources (detected with peak $\rm SNR \geq 5$), out of which 12 are multi-component. Size and flux density estimates from the high-resolution mosaic are reported as well.

  \item Monte Carlo simulations have been performed to derive the completeness of our master radio source catalog as a function of integrated flux densities. For flux densities larger than 10 (100)\,$\rm \mu Jy$, or $\rm SNR\approx 10$ (100) for unresolved sources, our catalog reaches a completeness of 50\% (100\%). The total fraction of spurious sources in our master catalog is only 0.75\%. 

  \item We have derived the 10\,GHz radio source counts in the GOODS-N field. Comparing our results with the 10\,GHz number counts from  a single-pointing VLA image in COSMOS \citep{vandervlugt21}, we find that the latter are systematically higher (by a factor $\approx 1.6$). This is likely a consequence of sample and/or cosmic variance arising from the small field of view of the VLA COSMOS observations.
  
  \item The 10\,GHz radio source counts in the GOODS-N field agree with the expected trend from  1.4\,GHz and 3\,GHz radio counts that are scaled to the 10\,GHz observed frame. Nevertheless, the number counts inferred from previous studies and this work scatter across $\rm \approx 0.2 - 0.3\,dex$, which might be driven by  the different approaches used to correct for observational biases, as well as field-to-field variations due to sample and cosmic variance. 

  \item The 10\,GHz radio source counts across ($ 16 \lesssim S_{\rm I} / \mu{\rm Jy} \lesssim 1600 $) are, in general, consistent with the predictions made by  \citet{mancuso17} and \citet{bonaldi19}. At the faint end ($\rm S_{\rm I}\approx 10\mu\, Jy$), the predictions from \citet{mancuso17} offer a better description of the 10\,GHz radio source counts in GOODS-N.

\end{itemize}

Since this is the deepest and most detailed radio survey of the high-frequency radio sky, the 10\,GHz mosaic of GOODS-N has the potential to address a diversity of open issues in extragalactic astronomy. The synthesized beam of $0\farcs22$ in the high-resolution mosaic approaches the angular resolution of {\it HST}  and {\it JWST}, allowing us to explore obscured and unobscured star formation in distant galaxies at similar angular resolutions. Combining 10\,GHz radio continuum imaging with available 1.4, 3, and 5\,GHz data in the GOODS-N field,  it will be possible to explore the radio spectro-morphological properties of $\mu \rm Jy$ sources  to benchmark radio continuum emission as a robust indicator of star formation at high redshifts. These studies will be reported in forthcoming manuscripts,  paving the way for the observational studies of sub-$\mu$Jy radio sources (at sub-arcsec resolutions) that will be routinely obtained with the next generation VLA in the next decade \citep[e.g.,][see Figure\,\ref{fig:surveys}]{barger18, murphy22, latif24}.


\begin{acknowledgments}
E.F.-J.A.  gratefully acknowledges the support and guidance to process this VLA Large Program from the NRAO IT  and the 
 Science Helpdesk teams, including  Frank Schinzel, Juergen Ott, Aaron Lawson, Drew Medlin,  K. Scott Rowe, Tracy Halstead, and Abi Smoake. E.F.-J.A. also  thanks Sanjay Bhatnagar and Preshanth Jagannathan for helpful discussions on the imaging  process. E.F.-J.A. acknowledge support from UNAM-PAPIIT project IA102023, and from CONAHCyT Ciencia de Frontera project ID:  CF-2023-I-506. 
The National Radio Astronomy Observatory is a facility of the National Science Foundation operated under cooperative agreement by Associated Universities, Inc. 
\end{acknowledgments}


\vspace{5mm}
\facilities{NSF's Karl G. Jansky Very Large Array (VLA)}


\software{Astropy \citep{2013A&A...558A..33A}, \citep{2018AJ....156..123A}, \citep{2022ApJ...935..167A}, APLpy \citep{2012ascl.soft08017R}, PyBDSF\citep{Mohan15}.}

%           Cloudy \citep{2013RMxAA..49..137F}, 
%           Source Extractor \citep{1996A&AS..117..393B}
%           }



\clearpage
\appendix
\restartappendixnumbering

\section{10\,GHz catalogs of GOODS-N}
We present a sample of the master 10\,GHz   catalog of GOODS-N in Table\,\ref{table:radiosourcecat}. It reports the  flux densities of sources measured in the low- and high-resolution mosaics. The radio size estimates from the high-resolution mosaic are provided as well if sources are simultaneously detected in the low- and high-resolution mosaics. Additionally, in Tables\,\ref{table:multi-component_cat_1d0} and \,\ref{table:multi-component_cat_0d2}, we present the information of the multiple islands of emission that constitute the multi-component radio sources as observed in the low- and high-resolution version of the mosaic, respectively.

\begin{longrotatetable}
\begin{deluxetable*}{l  l l r r r r r r r r r}
\tabletypesize{\ssmall}
\tablecolumns{12}
\tablecaption{ \footnotesize{A sample of the catalog of 256 radio sources in the 10\,GHz mosaic of the GOODS-N field at $1\farcs0$ resolution. The full catalog can be found in the online version of this manuscript and at \url{https://science.nrao.edu/science/surveys/vla-x-gn/home}. \label{table:radiosourcecat}}}
\tablehead{	\colhead{Name} & \colhead{RA$^\tablenotemark{a}$  }  & \colhead{DEC$^\tablenotemark{a}$}  &  \colhead{$S_{\rm I}^{\rm 1.0}$ $^\tablenotemark{b}$}  & \colhead{$S_{\rm P}^{\rm 1.0}$ $^\tablenotemark{c}$} & \colhead{$S_{\rm I}^{\rm 0.22}$ $^\tablenotemark{b}$} & \colhead{$S_{\rm P}^{\rm 0.22}$ $^\tablenotemark{c}$} & \colhead{$S_{*}^{\rm 0.22}$ $^\tablenotemark{d}$} &  \colhead{$\theta_{\rm M}$} & \colhead{$\theta_{\rm m}$} & \colhead{R$^\tablenotemark{e}$  } & \colhead{T$^\tablenotemark{f}$}   \\
\colhead{} & \colhead{(degs)}  & \colhead{(degs)}  &  \colhead{$(\rm \mu Jy)$}  & \colhead{$(\rm \mu Jy\, beam^{-1})$} & \colhead{$(\rm \mu Jy)$} & \colhead{$(\rm \mu Jy\, beam^{-1})$} & \colhead{$(\rm \mu Jy)$} &  \colhead{(mas)} & \colhead{(mas)} & \colhead{} & \colhead{} } 
\startdata
J123830.67+621821.42 & 189.627809 $\pm$ 0.000055 & 62.305950 $\pm$ 0.000031 & 170.80 $\pm$ 32.80 & 37.12 $\pm$ 5.94 &   &   &  &  &   &  & C \\
J123821.96+621823.83 & 189.591517 $\pm$ 0.000028 & 62.306619 $\pm$ 0.000015 & 29.86 $\pm$ 6.09 & 18.15 $\pm$ 2.46 & 20.15 $\pm$ 2.96 & 18.79 $\pm$ 1.63 & 19.46 $\pm$ 0.67 & 86 $\pm$ 58 & 0 $\pm$ 18 & 0 & C  \\
J123821.78+621706.68 & 189.590749 $\pm$ 0.000039 & 62.285190 $\pm$ 0.000025 & 22.28 $\pm$ 6.10 & 12.25 $\pm$ 2.29 &   &   &  &  &   &  & C \\
J123819.81+621839.22 & 189.582537 $\pm$ 0.000019 & 62.310894 $\pm$ 0.000025 & 55.38 $\pm$ 7.86 & 20.87 $\pm$ 2.23 &   &   &  &  &   &  & C \\
J123819.08+621826.99 & 189.579516 $\pm$ 0.000014 & 62.307498 $\pm$ 0.000010 & 15.31 $\pm$ 2.91 & 17.24 $\pm$ 1.81 & 23.69 $\pm$ 3.18 & 15.84 $\pm$ 1.38 & 23.69 $\pm$ 3.18 & 200 $\pm$ 42 & 104 $\pm$ 46 & 1 & C  \\
J123811.83+621821.90 & 189.549305 $\pm$ 0.000007 & 62.306084 $\pm$ 0.000007 & 31.59 $\pm$ 2.67 & 24.75 $\pm$ 1.31 & 23.54 $\pm$ 1.57 & 22.35 $\pm$ 0.88 & 22.94 $\pm$ 0.68 & 0 $\pm$ 27 & 0 $\pm$ 8 & 0 & C  \\
J123810.58+621729.23 & 189.544078 $\pm$ 0.000042 & 62.291454 $\pm$ 0.000016 & 7.69 $\pm$ 2.37 & 5.88 $\pm$ 1.11 & 8.10 $\pm$ 1.75 & 5.82 $\pm$ 0.79 & 6.87 $\pm$ 0.69 & 237 $\pm$ 72 & 0 $\pm$ 23 & 0 & C  \\
J123807.76+621540.57 & 189.532353 $\pm$ 0.000010 & 62.261270 $\pm$ 0.000010 & 15.16 $\pm$ 2.49 & 17.31 $\pm$ 1.57 & 26.82 $\pm$ 1.68 & 30.95 $\pm$ 1.07 & 28.81 $\pm$ 0.64 & 0 $\pm$ 8 & 0 $\pm$ 7 & 0 & C  \\
J123807.43+621650.69 & 189.530948 $\pm$ 0.000039 & 62.280746 $\pm$ 0.000032 & 14.58 $\pm$ 3.77 & 6.16 $\pm$ 1.17 &   &   &  &  &   &  & C \\
J123805.51+621445.55 & 189.522969 $\pm$ 0.000045 & 62.245986 $\pm$ 0.000039 & 31.26 $\pm$ 8.03 & 12.62 $\pm$ 2.38 &   &   &  &  &   &  & C \\
J123803.66+621711.37 & 189.515231 $\pm$ 0.000011 & 62.286491 $\pm$ 0.000031 & 43.15 $\pm$ 4.33 & 11.71 $\pm$ 0.94 &   &   &  &  &   &  & C \\
J123800.92+621336.02 & 189.503821 $\pm$ 0.000014 & 62.226672 $\pm$ 0.000020 & 25.10 $\pm$ 5.82 & 23.01 $\pm$ 3.15 & 42.29 $\pm$ 10.32 & 10.94 $\pm$ 2.17 & 42.29 $\pm$ 10.32 & 376 $\pm$ 100 & 369 $\pm$ 99 & 1 & C  \\
J123758.81+621458.32 & 189.495042 $\pm$ 0.000024 & 62.249533 $\pm$ 0.000021 & 19.17 $\pm$ 3.91 & 11.18 $\pm$ 1.54 & 13.99 $\pm$ 3.62 & 5.37 $\pm$ 1.04 & 13.99 $\pm$ 3.62 & 361 $\pm$ 106 & 201 $\pm$ 78 & 1 & C  \\
J123757.01+622059.41 & 189.487547 $\pm$ 0.000005 & 62.349835 $\pm$ 0.000005 & 139.62 $\pm$ 9.84 & 127.19 $\pm$ 5.33 & 136.81 $\pm$ 7.54 & 109.45 $\pm$ 3.74 & 136.81 $\pm$ 7.54 & 123 $\pm$ 18 & 97 $\pm$ 20 & 1 & C  \\
J123755.95+621507.54 & 189.483125 $\pm$ 0.000025 & 62.252093 $\pm$ 0.000035 & 19.78 $\pm$ 4.23 & 8.31 $\pm$ 1.30 & 7.45 $\pm$ 2.09 & 4.69 $\pm$ 0.84 & 5.91 $\pm$ 0.67 & 318 $\pm$ 110 & 0 $\pm$ 28 & 0 & C  \\
J123752.75+621628.25 & 189.469793 $\pm$ 0.000025 & 62.274515 $\pm$ 0.000023 & 17.07 $\pm$ 3.25 & 7.79 $\pm$ 1.07 & 15.59 $\pm$ 3.84 & 3.04 $\pm$ 0.63 & 15.59 $\pm$ 3.84 & 602 $\pm$ 155 & 318 $\pm$ 90 & 1 & C  \\
J123752.55+621936.96 & 189.468945 $\pm$ 0.000040 & 62.326933 $\pm$ 0.000048 & 23.39 $\pm$ 6.43 & 7.99 $\pm$ 1.69 &   &   &  &  &   &  & C \\
J123751.23+621919.02 & 189.463461 $\pm$ 0.000002 & 62.321949 $\pm$ 0.000002 & 118.56 $\pm$ 2.67 & 104.79 $\pm$ 1.42 & 109.07 $\pm$ 1.86 & 95.90 $\pm$ 0.99 & 109.07 $\pm$ 1.86 & 93 $\pm$ 6 & 68 $\pm$ 8 & 1 & C  \\
J123750.25+621359.13 & 189.459382 $\pm$ 0.000022 & 62.233091 $\pm$ 0.000031 & 12.27 $\pm$ 3.34 & 8.21 $\pm$ 1.45 & 6.80 $\pm$ 1.75 & 6.26 $\pm$ 0.95 & 6.52 $\pm$ 0.67 & 0 $\pm$ 83 & 0 $\pm$ 26 & 0 & C  \\
J123748.15+621610.37 & 189.450641 $\pm$ 0.000016 & 62.269546 $\pm$ 0.000023 & 7.79 $\pm$ 1.90 & 6.69 $\pm$ 0.98 & 7.99 $\pm$ 1.64 & 5.32 $\pm$ 0.71 & 6.52 $\pm$ 0.72 & 201 $\pm$ 64 & 105 $\pm$ 70 & 0 & C  \\
J123747.94+621442.10 & 189.449761 $\pm$ 0.000019 & 62.245027 $\pm$ 0.000017 & 10.33 $\pm$ 2.17 & 8.40 $\pm$ 1.08 & 7.93 $\pm$ 1.51 & 6.43 $\pm$ 0.75 & 7.15 $\pm$ 0.70 & 147 $\pm$ 60 & 49 $\pm$ 114 & 0 & C  \\
J123747.08+621631.90 & 189.446163 $\pm$ 0.000013 & 62.275528 $\pm$ 0.000014 & 12.43 $\pm$ 2.02 & 10.19 $\pm$ 1.02 & 15.09 $\pm$ 3.00 & 4.35 $\pm$ 0.69 & 15.09 $\pm$ 3.00 & 457 $\pm$ 98 & 247 $\pm$ 63 & 1 & C  \\
J123746.67+621738.59 & 189.444478 $\pm$ 0.000001 & 62.294053 $\pm$ 0.000001 & 311.87 $\pm$ 8.21 & 254.71 $\pm$ 1.03 & 258.37 $\pm$ 1.36 & 230.77 $\pm$ 0.73 & 258.37 $\pm$ 1.36 & 86 $\pm$ 2 & 65 $\pm$ 3 & 1 & C  \\
J123745.88+621434.83 & 189.441178 $\pm$ 0.000060 & 62.243010 $\pm$ 0.000079 & 55.52 $\pm$ 9.48 & 4.67 $\pm$ 0.74 &   &   &  &  &   &  & C \\
J123745.73+621456.55 & 189.440523 $\pm$ 0.000009 & 62.249043 $\pm$ 0.000009 & 20.30 $\pm$ 2.14 & 15.49 $\pm$ 1.03 & 16.86 $\pm$ 1.47 & 13.08 $\pm$ 0.71 & 16.86 $\pm$ 1.47 & 154 $\pm$ 27 & 75 $\pm$ 37 & 1 & C  \\
J123745.33+622023.75 & 189.438871 $\pm$ 0.000023 & 62.339931 $\pm$ 0.000053 & 19.11 $\pm$ 5.38 & 8.89 $\pm$ 1.78 & 10.04 $\pm$ 1.90 & 10.95 $\pm$ 1.15 & 10.48 $\pm$ 0.65 & 0 $\pm$ 88 & 0 $\pm$ 18 & 0 & C  \\
J123744.68+621218.76 & 189.436167 $\pm$ 0.000047 & 62.205212 $\pm$ 0.000044 & 57.75 $\pm$ 12.35 & 12.72 $\pm$ 2.27 & 20.58 $\pm$ 5.99 & 7.53 $\pm$ 1.66 & 20.58 $\pm$ 5.99 & 383 $\pm$ 125 & 203 $\pm$ 88 & 1 & C  \\
$\vdots$ & $\vdots$ & $\vdots$ & $\vdots$ & $\vdots$ & $\vdots$ & $\vdots$& $\vdots$ &  $\vdots$  & $\vdots$ & $\vdots$ & $\vdots$ \\
J123642.22+621545.48 & 189.175896 $\pm$ 0.000003 & 62.262634 $\pm$ 0.000003 & 54.77 $\pm$ 2.14 & 45.02 $\pm$ 1.08 & 47.62 $\pm$ 1.89 &   &   &  &  & &  M \\
J123629.01+621045.59 & 189.120878 $\pm$ 0.000026 & 62.179330 $\pm$ 0.000054 & 26.55 $\pm$ 4.71 & 7.29 $\pm$ 1.04 &   &  &  &  &  & & M \\
J123612.46+621140.48 & 189.051920 $\pm$ 0.000024 & 62.194578 $\pm$ 0.000027 & 26.28 $\pm$ 4.17 & 8.72 $\pm$ 1.07 & 9.63 $\pm$ 1.99 &   &   &  &  & &  M \\
J123531.57+621117.51 & 188.881551 $\pm$ 0.000023 & 62.188198 $\pm$ 0.000034 & 19.03 $\pm$ 4.53 & 10.70 $\pm$ 1.72 & 12.24 $\pm$ 2.71 &   &   &  &  & &  M \\
\enddata
\tablerefs{\footnotesize $^a$The rms position uncertainties are given by $\phi/[(2 \ln 2)^{1/2}\times \rm SNR$] \citep{condon98}.   $^b$$S_{\rm I}$ -- Integrated flux density. $^c$$S_{\rm P}$ -- Peak brightness.  $^d$$S_{\star}$ -- Best estimate for the integrated flux density (see Section\,\ref{subsec:radiosizeestimates}). $^e$If sources are confidently resolved along the major axis in the $0\farcs22$ resolution mosaic  ($\phi_{\rm M}-\theta_{\rm 1/2}\geq 2\sigma_{\rm \phi_{\rm M}}$), $\rm R=1$. Else, if $\phi_{\rm M}-\theta_{\rm 1/2} < 2\sigma_{\rm \phi_{\rm M}}$,  $\rm R=0$.   $^f$Source type. $\rm T=C$ for compact, single radio sources in the $1\farcs0$ resolution mosaic  and $\rm T=M$ for multi-component, extended and complex radio sources in the $1\farcs0$ or  $0\farcs22$ resolution mosaic (see Tables\,\ref{table:multi-component_cat_1d0} and \ref{table:multi-component_cat_0d2}).}
\end{deluxetable*}
\end{longrotatetable}

\begin{deluxetable*}{c  l l r r }
\tabletypesize{\footnotesize}
\tablecolumns{5}
\tablecaption{Flux densities and positions of the radio sources in the low-resolution ($1\farcs0$ resolution) mosaic that are part of the 12 multi-component radio sources in our master catalog.  \label{table:multi-component_cat_1d0}}
\tablehead{	\colhead{Name} & \colhead{RA$^{\rm 1.0}$$^\tablenotemark{a}$}  & \colhead{DEC$^{\rm 1.0}$$^\tablenotemark{a}$}  &  \colhead{$S_{\rm I}^{\rm 1.0}$ $^\tablenotemark{b}$}  & \colhead{$S_{\rm P}^{\rm 1.0}$ $^\tablenotemark{c}$}  \\
\colhead{} & \colhead{(degs)}  & \colhead{(degs)}  &  \colhead{$(\rm \mu Jy)$}  & \colhead{$(\rm \mu Jy\, beam^{-1})$}  }  
\startdata
J123820.47+621828.25  & 189.585301 $\pm$ 0.000006 & 62.307847 $\pm$ 0.000006 & 171.20 $\pm$ 7.71 & 86.55 $\pm$ 2.10  \\
\hline
J123725.92+621128.38  & 189.359143 $\pm$ 0.000017 & 62.191054 $\pm$ 0.000020 & 92.86 $\pm$ 7.49 & 18.83 $\pm$ 1.28  \\
  & 189.358146 $\pm$ 0.000002 & 62.191315 $\pm$ 0.000003 & 172.01 $\pm$ 3.83 & 99.40 $\pm$ 1.32  \\
  & 189.356857 $\pm$ 0.000010 & 62.191148 $\pm$ 0.000009 & 106.40 $\pm$ 5.81 & 31.50 $\pm$ 1.36  \\
\hline
J123717.90+621855.63  & 189.324579 $\pm$ 0.000020 & 62.315452 $\pm$ 0.000025 & 13.45 $\pm$ 2.77 & 7.83 $\pm$ 1.09  \\
\hline
J123711.96+621325.92  & 189.299983 $\pm$ 0.000017 & 62.223809 $\pm$ 0.000014 & 9.50 $\pm$ 1.95 & 8.39 $\pm$ 1.03  \\
  & 189.299380 $\pm$ 0.000046 & 62.224119 $\pm$ 0.000051 & 5.92 $\pm$ 2.56 & 3.72 $\pm$ 1.05  \\
\hline
J123711.31+621330.93  & 189.297126 $\pm$ 0.000029 & 62.225258 $\pm$ 0.000012 & 26.74 $\pm$ 3.58 & 10.54 $\pm$ 1.04  \\
\hline
J123707.99+621121.65  & 189.283196 $\pm$ 0.000024 & 62.189335 $\pm$ 0.000019 & 14.78 $\pm$ 2.80 & 8.00 $\pm$ 1.05  \\
  & 189.284158 $\pm$ 0.000042 & 62.189409 $\pm$ 0.000059 & 5.26 $\pm$ 2.42 & 3.37 $\pm$ 1.01  \\
\hline
J123645.81+620754.29  & 189.191129 $\pm$ 0.000023 & 62.131740 $\pm$ 0.000011 & 15.35 $\pm$ 2.54 & 9.77 $\pm$ 1.06  \\
 & 189.190348 $\pm$ 0.000016 & 62.131746 $\pm$ 0.000026 & 6.85 $\pm$ 1.88 & 6.29 $\pm$ 1.01  \\
\hline
J123644.40+621133.19  & 189.184950 $\pm$ 0.000001 & 62.192536 $\pm$ 0.000001 & 465.91 $\pm$ 3.21 & 402.52 $\pm$ 1.02  \\
 & 189.184883 $\pm$ 0.000012 & 62.191452 $\pm$ 0.000045 & 11.98 $\pm$ 2.70 & 6.83 $\pm$ 1.01  \\
\hline
J123642.22+621545.48  & 189.175896 $\pm$ 0.000003 & 62.262634 $\pm$ 0.000003 & 54.77 $\pm$ 2.14 & 45.02 $\pm$ 1.08  \\
\hline
J123629.01+621045.59  & 189.120427 $\pm$ 0.000017 & 62.179271 $\pm$ 0.000026 & 10.71 $\pm$ 2.40 & 7.26 $\pm$ 1.05  \\
 & 189.121352 $\pm$ 0.000037 & 62.179412 $\pm$ 0.000049 & 17.31 $\pm$ 4.14 & 5.45 $\pm$ 1.02  \\
\hline
J123612.46+621140.48  & 189.051920 $\pm$ 0.000024 & 62.194578 $\pm$ 0.000027 & 26.28 $\pm$ 4.17 & 8.72 $\pm$ 1.07  \\
\hline
J123531.57+621117.51  & 188.881551 $\pm$ 0.000023 & 62.188198 $\pm$ 0.000034 & 19.03 $\pm$ 4.53 & 10.70 $\pm$ 1.72  \\
\enddata
\tablerefs{\footnotesize $^a$The rms position uncertainties are given by $\phi/[(2 \ln 2)^{1/2}\times \rm SNR$] \citep{condon98}. $^b$$S_{\rm I}$ -- Integrated flux density. $^c$$S_{\rm P}$ -- Peak brightness. }
\end{deluxetable*}

\begin{longrotatetable}
\begin{deluxetable*}{c  l l r r r r r  r  }
\tabletypesize{\footnotesize}
\tablecolumns{9}
\tablecaption{Flux densities, sizes, and positions of the radio sources in the $0\farcs22$ resolution mosaic that are part of the 11 multi-component radio sources in our master catalog that have a counterpart in the high-resolution map.  \label{table:multi-component_cat_0d2}}
\tablehead{	\colhead{Name} & \colhead{RA$^{\rm 0.22}$$^\tablenotemark{a}$}  & \colhead{DEC$^{\rm 0.22}$$^\tablenotemark{a}$}  &  \colhead{$S_{\rm I}^{\rm 0.22}$ $^\tablenotemark{b}$}  & \colhead{$S_{\rm P}^{\rm 0.22}$ $^\tablenotemark{c}$} & \colhead{$S_{*}^{\rm 0.22}$ $^\tablenotemark{d}$} &  \colhead{$\theta_{\rm M}$} & \colhead{$\theta_{\rm m}$} & \colhead{R$^\tablenotemark{e}$  }   \\
\colhead{} & \colhead{(degs)}  & \colhead{(degs)}  &  \colhead{$(\rm \mu Jy)$} &   \colhead{$(\rm \mu Jy\, beam^{-1})$}  & \colhead{$(\rm \mu Jy)$}  & \colhead{(mas)} & \colhead{(mas)} & \colhead{} }  
\startdata
J123820.47+621828.25  & 189.585226 $\pm$ 0.000012 & 62.307875 $\pm$ 0.000011 & 196.28 $\pm$ 16.97 & 30.03 $\pm$ 1.45 & 196.28 $\pm$ 16.97 & 1169 $\pm$ 140 & 167 $\pm$ 33 & 1  \\
  & 189.584946 $\pm$ 0.000004 & 62.308504 $\pm$ 0.000005 & 12.04 $\pm$ 3.03 & 9.80 $\pm$ 1.52 & 10.86 $\pm$ 0.69 & 132 $\pm$ 81 & 73 $\pm$ 109 & 0  \\
\hline
J123725.92+621128.38  & 189.358115 $\pm$ 0.000001 & 62.191306 $\pm$ 0.000001 & 38.35 $\pm$ 1.79 & 32.74 $\pm$ 0.93 & 38.35 $\pm$ 1.79 & 106 $\pm$ 16 & 74 $\pm$ 20 & 1  \\
  & 189.358420 $\pm$ 0.000001 & 62.191346 $\pm$ 0.000002 & 71.16 $\pm$ 3.34 & 31.70 $\pm$ 0.90 & 71.16 $\pm$ 3.34 & 311 $\pm$ 18 & 125 $\pm$ 15 & 1  \\
  & 189.357841 $\pm$ 0.000001 & 62.191286 $\pm$ 0.000001 & 55.16 $\pm$ 2.84 & 26.40 $\pm$ 0.97 & 55.16 $\pm$ 2.84 & 251 $\pm$ 17 & 208 $\pm$ 16 & 1  \\
\hline
J123717.90+621855.63  & 189.324547 $\pm$ 0.000006 & 62.315484 $\pm$ 0.000004 & 4.29 $\pm$ 1.30 & 3.98 $\pm$ 0.70 & 4.13 $\pm$ 0.67 & 0 $\pm$ 97 & 0 $\pm$ 29 & 0  \\
  & 189.324573 $\pm$ 0.000005 & 62.315406 $\pm$ 0.000010 & 5.36 $\pm$ 1.78 & 3.38 $\pm$ 0.73 & 4.26 $\pm$ 0.70 & 272 $\pm$ 117 & 0 $\pm$ 40 & 0  \\
   \hline
J123711.96+621325.92  & 189.299945 $\pm$ 0.000003 & 62.223830 $\pm$ 0.000004 & 8.90 $\pm$ 1.55 & 6.78 $\pm$ 0.74 & 7.77 $\pm$ 0.72 & 183 $\pm$ 54 & 33 $\pm$ 148 & 0  \\
  \hline
J123711.31+621330.93  & 189.297436 $\pm$ 0.000004 & 62.225281 $\pm$ 0.000005 & 6.94 $\pm$ 1.61 & 5.13 $\pm$ 0.75 & 5.97 $\pm$ 0.73 & 163 $\pm$ 72 & 94 $\pm$ 84 & 0  \\
  & 189.296870 $\pm$ 0.000002 & 62.225240 $\pm$ 0.000002 & 8.02 $\pm$ 1.16 & 8.63 $\pm$ 0.70 & 8.32 $\pm$ 0.67 & 0 $\pm$ 80 & 0 $\pm$ 15 & 0  \\
   \hline
J123707.99+621121.65  & 189.283196 $\pm$ 0.000008 & 62.189312 $\pm$ 0.000006 & 8.26 $\pm$ 2.11 & 3.94 $\pm$ 0.72 & 8.26 $\pm$ 2.11 & 293 $\pm$ 91 & 169 $\pm$ 76 & 1  \\
\hline
J123645.81+620754.29  & 189.191093 $\pm$ 0.000015 & 62.131730 $\pm$ 0.000007 & 12.33 $\pm$ 3.17 & 3.40 $\pm$ 0.70 & 12.33 $\pm$ 3.17 & 502 $\pm$ 138 & 233 $\pm$ 80 & 1  \\
  & 189.190412 $\pm$ 0.000009 & 62.131725 $\pm$ 0.000011 & 12.40 $\pm$ 2.88 & 3.93 $\pm$ 0.71 & 12.40 $\pm$ 2.88 & 466 $\pm$ 118 & 198 $\pm$ 70 & 1  \\
\hline
J123644.40+621133.19  & 189.184941 $\pm$ 0.000001 & 62.192538 $\pm$ 0.000001 & 414.04 $\pm$ 2.82 & 339.97 $\pm$ 0.70 & 414.04 $\pm$ 2.82 & 250 $\pm$ 1 & 66 $\pm$ 2 & 1  \\
\hline
J123642.22+621545.48  & 189.175882 $\pm$ 0.000001 & 62.262644 $\pm$ 0.000001 & 44.68 $\pm$ 1.33 & 41.63 $\pm$ 0.73 & 44.68 $\pm$ 1.33 & 78 $\pm$ 13 & 34 $\pm$ 25 & 1  \\
  & 189.175916 $\pm$ 0.000010 & 62.262546 $\pm$ 0.000006 & 2.89 $\pm$ 1.34 & 2.66 $\pm$ 0.72 & 2.77 $\pm$ 0.69 & 0 $\pm$ 148 & 0 $\pm$ 47 & 0  \\
\hline
J123612.46+621140.48  & 189.051956 $\pm$ 0.000004 & 62.194610 $\pm$ 0.000004 & 6.69 $\pm$ 1.49 & 5.37 $\pm$ 0.74 & 5.99 $\pm$ 0.71 & 132 $\pm$ 72 & 82 $\pm$ 89 & 0  \\
  & 189.051860 $\pm$ 0.000006 & 62.194679 $\pm$ 0.000013 & 2.95 $\pm$ 1.33 & 2.63 $\pm$ 0.67 & 2.79 $\pm$ 0.64 & 0 $\pm$ 155 & 0 $\pm$ 31 & 0  \\
\hline
J123531.57+621117.51  & 188.881592 $\pm$ 0.000004 & 62.188180 $\pm$ 0.000005 & 5.60 $\pm$ 1.82 & 6.13 $\pm$ 1.11 & 5.86 $\pm$ 0.65 & 0 $\pm$ 312 & 0 $\pm$ 34 & 0  \\
  & 188.881497 $\pm$ 0.000006 & 62.188251 $\pm$ 0.000005 & 6.64 $\pm$ 2.02 & 6.46 $\pm$ 1.10 & 6.55 $\pm$ 0.65 & 175 $\pm$ 97 & 0 $\pm$ 24 & 0  \\
\enddata
\tablerefs{$^a$The rms position uncertainties are given by $\phi/[(2 \ln 2)^{1/2}\times \rm SNR$] \citep{condon98}. $^b$$S_{\rm I}$ -- Integrated flux density. $^c$$S_{\rm P}$ -- Peak brightness.  $^d$$S_{\star}$ -- Best estimate for the integrated flux density (see Section\,\ref{subsec:radiosizeestimates}). $^e$If sources are confidently resolved along the major axis in the $0\farcs22$ resolution mosaic  ($\phi_{\rm M}-\theta_{\rm 1/2}\geq 2\sigma_{\rm \phi_{\rm M}}$), $\rm R=1$. Else, if $\phi_{\rm M}-\theta_{\rm 1/2} < 2\sigma_{\rm \phi_{\rm M}}$,  $\rm R=0$. }
\end{deluxetable*}
\end{longrotatetable}


\bibliography{jimenezandrade+24}{}

\begin{thebibliography}{}
\expandafter\ifx\csname natexlab\endcsname\relax\def\natexlab#1{#1}\fi
\providecommand{\url}[1]{\href{#1}{#1}}
\providecommand{\dodoi}[1]{doi:~\href{http://doi.org/#1}{\nolinkurl{#1}}}
\providecommand{\doeprint}[1]{\href{http://ascl.net/#1}{\nolinkurl{http://ascl.net/#1}}}
\providecommand{\doarXiv}[1]{\href{https://arxiv.org/abs/#1}{\nolinkurl{https://arxiv.org/abs/#1}}}

\bibitem[{{Afonso} {et~al.}(2001){Afonso}, {Mobasher}, {Hopkins}, \&
  {Cram}}]{afonso01}
{Afonso}, J., {Mobasher}, B., {Hopkins}, A., \& {Cram}, L. 2001, \apss, 276,
  941, \dodoi{10.1023/A:1017578920588}

\bibitem[{{Algera} {et~al.}(2021){Algera}, {Hodge}, {Riechers}, {Murphy},
  {Pavesi}, {Aravena}, {Daddi}, {Decarli}, {Dickinson}, {Sargent}, {Sharon}, \&
  {Wagg}}]{algera21}
{Algera}, H.~S.~B., {Hodge}, J.~A., {Riechers}, D., {et~al.} 2021, \apj, 912,
  73, \dodoi{10.3847/1538-4357/abe6a5}

\bibitem[{Algera {et~al.}(2022)Algera, Hodge, Riechers, Leslie, Smail, Aravena,
  da~Cunha, Daddi, Decarli, Dickinson, Gim, Guaita, Magnelli, Murphy, Pavesi,
  Sargent, Sharon, Wagg, Walter, \& Yun}]{algera_22}
Algera, H. S.~B., Hodge, J.~A., Riechers, D.~A., {et~al.} 2022, The
  Astrophysical Journal, 924, 76, \dodoi{10.3847/1538-4357/ac34f5}

\bibitem[{{Amarantidis} {et~al.}(2023){Amarantidis}, {Afonso}, {Matute},
  {Farrah}, {Hopkins}, {Messias}, {Pappalardo}, \& {Seymour}}]{amarantidis23}
{Amarantidis}, S., {Afonso}, J., {Matute}, I., {et~al.} 2023, \aap, 678, A116,
  \dodoi{10.1051/0004-6361/202346411}

\bibitem[{An {et~al.}(2024)An, Vaccari, Best, Ocran, Ishwara-Chandra, Taylor,
  Leslie, R{\"o}ttgering, Kondapally, Haskell, Collier, \& Bonato}]{An24}
An, F., Vaccari, M., Best, P.~N., {et~al.} 2024, Monthly Notices of the Royal
  Astronomical Society, 528, 5346, \dodoi{10.1093/mnras/stae364}

\bibitem[{{Astropy Collaboration} {et~al.}(2013){Astropy Collaboration},
  {Robitaille}, {Tollerud}, {Greenfield}, {Droettboom}, {Bray}, {Aldcroft},
  {Davis}, {Ginsburg}, {Price-Whelan}, {Kerzendorf}, {Conley}, {Crighton},
  {Barbary}, {Muna}, {Ferguson}, {Grollier}, {Parikh}, {Nair}, {Unther},
  {Deil}, {Woillez}, {Conseil}, {Kramer}, {Turner}, {Singer}, {Fox}, {Weaver},
  {Zabalza}, {Edwards}, {Azalee Bostroem}, {Burke}, {Casey}, {Crawford},
  {Dencheva}, {Ely}, {Jenness}, {Labrie}, {Lim}, {Pierfederici}, {Pontzen},
  {Ptak}, {Refsdal}, {Servillat}, \& {Streicher}}]{2013A&A...558A..33A}
{Astropy Collaboration}, {Robitaille}, T.~P., {Tollerud}, E.~J., {et~al.} 2013,
  \aap, 558, A33, \dodoi{10.1051/0004-6361/201322068}

\bibitem[{{Astropy Collaboration} {et~al.}(2018){Astropy Collaboration},
  {Price-Whelan}, {Sip{\H{o}}cz}, {G{\"u}nther}, {Lim}, {Crawford}, {Conseil},
  {Shupe}, {Craig}, {Dencheva}, {Ginsburg}, {VanderPlas}, {Bradley},
  {P{\'e}rez-Su{\'a}rez}, {de Val-Borro}, {Aldcroft}, {Cruz}, {Robitaille},
  {Tollerud}, {Ardelean}, {Babej}, {Bach}, {Bachetti}, {Bakanov}, {Bamford},
  {Barentsen}, {Barmby}, {Baumbach}, {Berry}, {Biscani}, {Boquien}, {Bostroem},
  {Bouma}, {Brammer}, {Bray}, {Breytenbach}, {Buddelmeijer}, {Burke},
  {Calderone}, {Cano Rodr{\'\i}guez}, {Cara}, {Cardoso}, {Cheedella}, {Copin},
  {Corrales}, {Crichton}, {D'Avella}, {Deil}, {Depagne}, {Dietrich}, {Donath},
  {Droettboom}, {Earl}, {Erben}, {Fabbro}, {Ferreira}, {Finethy}, {Fox},
  {Garrison}, {Gibbons}, {Goldstein}, {Gommers}, {Greco}, {Greenfield},
  {Groener}, {Grollier}, {Hagen}, {Hirst}, {Homeier}, {Horton}, {Hosseinzadeh},
  {Hu}, {Hunkeler}, {Ivezi{\'c}}, {Jain}, {Jenness}, {Kanarek}, {Kendrew},
  {Kern}, {Kerzendorf}, {Khvalko}, {King}, {Kirkby}, {Kulkarni}, {Kumar},
  {Lee}, {Lenz}, {Littlefair}, {Ma}, {Macleod}, {Mastropietro}, {McCully},
  {Montagnac}, {Morris}, {Mueller}, {Mumford}, {Muna}, {Murphy}, {Nelson},
  {Nguyen}, {Ninan}, {N{\"o}the}, {Ogaz}, {Oh}, {Parejko}, {Parley}, {Pascual},
  {Patil}, {Patil}, {Plunkett}, {Prochaska}, {Rastogi}, {Reddy Janga},
  {Sabater}, {Sakurikar}, {Seifert}, {Sherbert}, {Sherwood-Taylor}, {Shih},
  {Sick}, {Silbiger}, {Singanamalla}, {Singer}, {Sladen}, {Sooley},
  {Sornarajah}, {Streicher}, {Teuben}, {Thomas}, {Tremblay}, {Turner},
  {Terr{\'o}n}, {van Kerkwijk}, {de la Vega}, {Watkins}, {Weaver}, {Whitmore},
  {Woillez}, {Zabalza}, \& {Astropy Contributors}}]{2018AJ....156..123A}
{Astropy Collaboration}, {Price-Whelan}, A.~M., {Sip{\H{o}}cz}, B.~M., {et~al.}
  2018, \aj, 156, 123, \dodoi{10.3847/1538-3881/aabc4f}

\bibitem[{{Astropy Collaboration} {et~al.}(2022){Astropy Collaboration},
  {Price-Whelan}, {Lim}, {Earl}, {Starkman}, {Bradley}, {Shupe}, {Patil},
  {Corrales}, {Brasseur}, {N{\"o}the}, {Donath}, {Tollerud}, {Morris},
  {Ginsburg}, {Vaher}, {Weaver}, {Tocknell}, {Jamieson}, {van Kerkwijk},
  {Robitaille}, {Merry}, {Bachetti}, {G{\"u}nther}, {Aldcroft},
  {Alvarado-Montes}, {Archibald}, {B{\'o}di}, {Bapat}, {Barentsen},
  {Baz{\'a}n}, {Biswas}, {Boquien}, {Burke}, {Cara}, {Cara}, {Conroy},
  {Conseil}, {Craig}, {Cross}, {Cruz}, {D'Eugenio}, {Dencheva}, {Devillepoix},
  {Dietrich}, {Eigenbrot}, {Erben}, {Ferreira}, {Foreman-Mackey}, {Fox},
  {Freij}, {Garg}, {Geda}, {Glattly}, {Gondhalekar}, {Gordon}, {Grant},
  {Greenfield}, {Groener}, {Guest}, {Gurovich}, {Handberg}, {Hart},
  {Hatfield-Dodds}, {Homeier}, {Hosseinzadeh}, {Jenness}, {Jones}, {Joseph},
  {Kalmbach}, {Karamehmetoglu}, {Ka{\l}uszy{\'n}ski}, {Kelley}, {Kern},
  {Kerzendorf}, {Koch}, {Kulumani}, {Lee}, {Ly}, {Ma}, {MacBride}, {Maljaars},
  {Muna}, {Murphy}, {Norman}, {O'Steen}, {Oman}, {Pacifici}, {Pascual},
  {Pascual-Granado}, {Patil}, {Perren}, {Pickering}, {Rastogi}, {Roulston},
  {Ryan}, {Rykoff}, {Sabater}, {Sakurikar}, {Salgado}, {Sanghi}, {Saunders},
  {Savchenko}, {Schwardt}, {Seifert-Eckert}, {Shih}, {Jain}, {Shukla}, {Sick},
  {Simpson}, {Singanamalla}, {Singer}, {Singhal}, {Sinha}, {Sip{\H{o}}cz},
  {Spitler}, {Stansby}, {Streicher}, {{\v{S}}umak}, {Swinbank}, {Taranu},
  {Tewary}, {Tremblay}, {de Val-Borro}, {Van Kooten}, {Vasovi{\'c}}, {Verma},
  {de Miranda Cardoso}, {Williams}, {Wilson}, {Winkel}, {Wood-Vasey}, {Xue},
  {Yoachim}, {Zhang}, {Zonca}, \& {Astropy Project
  Contributors}}]{2022ApJ...935..167A}
{Astropy Collaboration}, {Price-Whelan}, A.~M., {Lim}, P.~L., {et~al.} 2022,
  \apj, 935, 167, \dodoi{10.3847/1538-4357/ac7c74}

\bibitem[{{Barger} {et~al.}(2018){Barger}, {Kohno}, {Murphy}, {Sargent}, \&
  {Condon}}]{barger18}
{Barger}, A.~J., {Kohno}, K., {Murphy}, E.~J., {Sargent}, M.~T., \& {Condon},
  J.~J. 2018, arXiv e-prints, arXiv:1810.07143,
  \dodoi{10.48550/arXiv.1810.07143}

\bibitem[{{Barro} {et~al.}(2019){Barro}, {P{\'e}rez-Gonz{\'a}lez}, {Cava},
  {Brammer}, {Pandya}, {Eliche Moral}, {Esquej}, {Dom{\'\i}nguez-S{\'a}nchez},
  {Alcalde Pampliega}, {Guo}, {Koekemoer}, {Trump}, {Ashby}, {Cardiel},
  {Castellano}, {Conselice}, {Dickinson}, {Dolch}, {Donley}, {Espino Briones},
  {Faber}, {Fazio}, {Ferguson}, {Finkelstein}, {Fontana}, {Galametz},
  {Gardner}, {Gawiser}, {Giavalisco}, {Grazian}, {Grogin}, {Hathi}, {Hemmati},
  {Hern{\'a}n-Caballero}, {Kocevski}, {Koo}, {Kodra}, {Lee}, {Lin}, {Lucas},
  {Mobasher}, {McGrath}, {Nandra}, {Nayyeri}, {Newman}, {Pforr}, {Peth},
  {Rafelski}, {Rodr{\'\i}guez-Munoz}, {Salvato}, {Stefanon}, {van der Wel},
  {Willner}, {Wiklind}, \& {Wuyts}}]{barro19}
{Barro}, G., {P{\'e}rez-Gonz{\'a}lez}, P.~G., {Cava}, A., {et~al.} 2019, \apjs,
  243, 22, \dodoi{10.3847/1538-4365/ab23f2}

\bibitem[{{Baugh} {et~al.}(2019){Baugh}, {Gonzalez-Perez}, {Lagos}, {Lacey},
  {Helly}, {Jenkins}, {Frenk}, {Benson}, {Bower}, \& {Cole}}]{baugh19}
{Baugh}, C.~M., {Gonzalez-Perez}, V., {Lagos}, C. D.~P., {et~al.} 2019, \mnras,
  483, 4922, \dodoi{10.1093/mnras/sty3427}

\bibitem[{{Best} {et~al.}(2023){Best}, {Kondapally}, {Williams}, {Cochrane},
  {Duncan}, {Hale}, {Haskell}, {Ma{\l}ek}, {McCheyne}, {Smith}, {Wang},
  {Botteon}, {Bonato}, {Bondi}, {Calistro Rivera}, {Gao}, {G{\"u}rkan},
  {Hardcastle}, {Jarvis}, {Mingo}, {Miraghaei}, {Morabito}, {Nisbet},
  {Prandoni}, {R{\"o}ttgering}, {Sabater}, {Shimwell}, {Tasse}, \& {van
  Weeren}}]{best23}
{Best}, P.~N., {Kondapally}, R., {Williams}, W.~L., {et~al.} 2023, \mnras, 523,
  1729, \dodoi{10.1093/mnras/stad1308}

\bibitem[{{Bhatnagar} {et~al.}(2013){Bhatnagar}, {Rau}, \&
  {Golap}}]{bhatnagar13}
{Bhatnagar}, S., {Rau}, U., \& {Golap}, K. 2013, \apj, 770, 91,
  \dodoi{10.1088/0004-637X/770/2/91}

\bibitem[{{Biggs} \& {Ivison}(2008)}]{bigss08}
{Biggs}, A.~D., \& {Ivison}, R.~J. 2008, \mnras, 385, 893,
  \dodoi{10.1111/j.1365-2966.2008.12869.x}

\bibitem[{{Bolton} {et~al.}(2004){Bolton}, {Cotter}, {Pooley}, {Riley},
  {Waldram}, {Chandler}, {Mason}, {Pearson}, \& {Readhead}}]{bolton04}
{Bolton}, R.~C., {Cotter}, G., {Pooley}, G.~G., {et~al.} 2004, \mnras, 354,
  485, \dodoi{10.1111/j.1365-2966.2004.08207.x}

\bibitem[{{Bonaldi} {et~al.}(2019){Bonaldi}, {Bonato}, {Galluzzi}, {Harrison},
  {Massardi}, {Kay}, {De Zotti}, \& {Brown}}]{bonaldi19}
{Bonaldi}, A., {Bonato}, M., {Galluzzi}, V., {et~al.} 2019, \mnras, 482, 2,
  \dodoi{10.1093/mnras/sty2603}

\bibitem[{Bondi {et~al.}(2018)Bondi, {Zamorani, G.}, {Ciliegi, P.},
  {Smolci\'{}c, V.}, {Schinnerer, E.}, {Delvecchio, I.}, {Jim\'enez-Andrade, E.
  F.}, {Liu, D.}, {Lang, P.}, {Magnelli, B.}, {Murphy, E. J.}, \& {Vardoulaki,
  E.}}]{bondi18}
Bondi, M., {Zamorani, G.}, {Ciliegi, P.}, {et~al.} 2018, A\&A, 618, L8,
  \dodoi{10.1051/0004-6361/201834243}

\bibitem[{{CASA Team} {et~al.}(2022){CASA Team}, Bean, Bhatnagar, Castro,
  Meyer, Emonts, Garcia, Garwood, Golap, Villalba, Harris, Hayashi, Hoskins,
  Hsieh, Jagannathan, Kawasaki, Keimpema, Kettenis, Lopez, Marvil, Masters,
  McNichols, Mehringer, Miel, Moellenbrock, Montesino, Nakazato, Ott, Petry,
  Pokorny, Raba, Rau, Schiebel, Schweighart, Sekhar, Shimada, Small, Steeb,
  Sugimoto, Suoranta, Tsutsumi, van Bemmel, Verkouter, Wells, Xiong, Szomoru,
  Griffith, Glendenning, \& Kern}]{casateam22}
{CASA Team}, T., Bean, B., Bhatnagar, S., {et~al.} 2022, Publications of the
  Astronomical Society of the Pacific, 134, 114501,
  \dodoi{10.1088/1538-3873/ac9642}

\bibitem[{{Casey} {et~al.}(2014){Casey}, {Narayanan}, \& {Cooray}}]{casey14}
{Casey}, C.~M., {Narayanan}, D., \& {Cooray}, A. 2014, Phys. Rev, 541, 45,
  \dodoi{10.1016/j.physrep.2014.02.009}

\bibitem[{{Condon}(1992)}]{condon92}
{Condon}, J.~J. 1992, ARA\&A, 30, 575,
  \dodoi{10.1146/annurev.aa.30.090192.003043}

\bibitem[{{Condon} {et~al.}(1998){Condon}, {Cotton}, {Greisen}, {Yin},
  {Perley}, {Taylor}, \& {Broderick}}]{condon98}
{Condon}, J.~J., {Cotton}, W.~D., {Greisen}, E.~W., {et~al.} 1998, \aj, 115,
  1693, \dodoi{10.1086/300337}

\bibitem[{Coppin {et~al.}(2005)Coppin, Halpern, {Scott}, {Borys}, \&
  {Chapman}}]{coppin05}
Coppin, K., Halpern, M., {Scott}, D., {Borys}, C., \& {Chapman}, S. 2005,
  MNRAS, 357, 1022, \dodoi{10.1111/j.1365-2966.2005.08723.x}

\bibitem[{Cotton {et~al.}(2018)Cotton, Condon, Kellermann, Lacy, Perley,
  Matthews, Vernstrom, Scott, \& Wall}]{cotton18}
Cotton, W.~D., Condon, J.~J., Kellermann, K.~I., {et~al.} 2018, \apj, 856, 67,
  \dodoi{10.3847/1538-4357/aaaec4}

\bibitem[{{de Zotti} {et~al.}(2005){de Zotti}, {Ricci}, {Mesa}, {Silva},
  {Mazzotta}, {Toffolatti}, \& {Gonz{\'a}lez-Nuevo}}]{dezotti05}
{de Zotti}, G., {Ricci}, R., {Mesa}, D., {et~al.} 2005, \aap, 431, 893,
  \dodoi{10.1051/0004-6361:20042108}

\bibitem[{Decarli {et~al.}(2020)Decarli, Aravena, Boogaard, Carilli,
  Gonz{\'a}lez-L{\'o}pez, Walter, Cortes, Cox, Cunha, Daddi, D{\'\i}az-Santos,
  Hodge, Inami, Neeleman, Novak, Oesch, Popping, Riechers, Smail, Uzgil, Werf,
  Wagg, \& Weiss}]{decarli20}
Decarli, R., Aravena, M., Boogaard, L., {et~al.} 2020, The Astrophysical
  Journal, 902, 110, \dodoi{10.3847/1538-4357/abaa3b}

\bibitem[{{Delvecchio} {et~al.}(2017){Delvecchio}, {Smol{\v c}i{\'c}, V.},
  {Zamorani, G.}, {Lagos, C. Del P.}, {Berta, S.}, {Delhaize, J.}, {Baran, N.},
  {Alexander, D. M.}, {Rosario, D. J.}, {Gonzalez-Perez, V.}, {Ilbert, O.},
  {Lacey, C. G.}, {Le F{\`e}vre, O.}, {Miettinen, O.}, {Aravena, M.}, {Bondi,
  M.}, {Carilli, C.}, {Ciliegi, P.}, {Mooley, K.}, {Novak, M.}, {Schinnerer,
  E.}, {Capak, P.}, {Civano, F.}, {Fanidakis, N.}, {Herrera Ruiz, N.}, {Karim,
  A.}, {Laigle, C.}, {Marchesi, S.}, {McCracken, H. J.}, {Middleberg, E.},
  {Salvato, M.}, \& {Tasca, L.}}]{delvecchio17}
{Delvecchio}, I., {Smol{\v c}i{\'c}, V.}, {Zamorani, G.}, {et~al.} 2017, A\&A,
  602, A3, \dodoi{10.1051/0004-6361/201629367}

\bibitem[{{Dickinson} {et~al.}(2003){Dickinson}, {Giavalisco}, \& {GOODS
  Team}}]{dickinson03}
{Dickinson}, M., {Giavalisco}, M., \& {GOODS Team}. 2003, in The Mass of
  Galaxies at Low and High Redshift, ed. R.~{Bender} \& A.~{Renzini}, 324,
  \dodoi{10.1007/10899892_78}

\bibitem[{{Driver} \& {Robotham}(2010)}]{driver10}
{Driver}, S.~P., \& {Robotham}, A. S.~G. 2010, \mnras, 407, 2131,
  \dodoi{10.1111/j.1365-2966.2010.17028.x}

\bibitem[{Dubner \& Giacani(2015)}]{dubner15}
Dubner, G., \& Giacani, E. 2015, The Astronomy and Astrophysics Review, 23, 3,
  \dodoi{10.1007/s00159-015-0083-5}

\bibitem[{{Eisenstein} {et~al.}(2023){Eisenstein}, {Willott}, {Alberts},
  {Arribas}, {Bonaventura}, {Bunker}, {Cameron}, {Carniani}, {Charlot},
  {Curtis-Lake}, {D'Eugenio}, {Endsley}, {Ferruit}, {Giardino}, {Hainline},
  {Hausen}, {Jakobsen}, {Johnson}, {Maiolino}, {Rieke}, {Rieke}, {Rix},
  {Robertson}, {Stark}, {Tacchella}, {Williams}, {Willmer}, {Baker}, {Baum},
  {Bhatawdekar}, {Boyett}, {Chen}, {Chevallard}, {Circosta}, {Curti},
  {Danhaive}, {DeCoursey}, {de Graaff}, {Dressler}, {Egami}, {Helton},
  {Hviding}, {Ji}, {Jones}, {Kumari}, {L{\"u}tzgendorf}, {Laseter}, {Looser},
  {Lyu}, {Maseda}, {Nelson}, {Parlanti}, {Perna}, {Pusk{\'a}s}, {Rawle},
  {Rodr{\'\i}guez Del Pino}, {Sandles}, {Saxena}, {Scholtz}, {Sharpe},
  {Shivaei}, {Silcock}, {Simmonds}, {Skarbinski}, {Smit}, {Stone}, {Suess},
  {Sun}, {Tang}, {Topping}, {{\"U}bler}, {Villanueva}, {Wallace}, {Whitler},
  {Witstok}, \& {Woodrum}}]{eisenstein23}
{Eisenstein}, D.~J., {Willott}, C., {Alberts}, S., {et~al.} 2023, arXiv
  e-prints, arXiv:2306.02465, \dodoi{10.48550/arXiv.2306.02465}

\bibitem[{{Fomalont} {et~al.}(2002){Fomalont}, {Kellermann}, {Partridge},
  {Windhorst}, \& {Richards}}]{fomalont02}
{Fomalont}, E.~B., {Kellermann}, K.~I., {Partridge}, R.~B., {Windhorst}, R.~A.,
  \& {Richards}, E.~A. 2002, \aj, 123, 2402, \dodoi{10.1086/339308}

\bibitem[{{Giavalisco} {et~al.}(2004){Giavalisco}, {Ferguson}, {Koekemoer},
  {Dickinson}, {Alexander}, {Bauer}, {Bergeron}, {Biagetti}, {Brandt},
  {Casertano}, {Cesarsky}, {Chatzichristou}, {Conselice}, {Cristiani}, {Da
  Costa}, {Dahlen}, {de Mello}, {Eisenhardt}, {Erben}, {Fall}, {Fassnacht},
  {Fosbury}, {Fruchter}, {Gardner}, {Grogin}, {Hook}, {Hornschemeier}, {Idzi},
  {Jogee}, {Kretchmer}, {Laidler}, {Lee}, {Livio}, {Lucas}, {Madau},
  {Mobasher}, {Moustakas}, {Nonino}, {Padovani}, {Papovich}, {Park},
  {Ravindranath}, {Renzini}, {Richardson}, {Riess}, {Rosati}, {Schirmer},
  {Schreier}, {Somerville}, {Spinrad}, {Stern}, {Stiavelli}, {Strolger},
  {Urry}, {Vandame}, {Williams}, \& {Wolf}}]{giavalisco04}
{Giavalisco}, M., {Ferguson}, H.~C., {Koekemoer}, A.~M., {et~al.} 2004, \apjl,
  600, L93, \dodoi{10.1086/379232}

\bibitem[{{Gim} {et~al.}(2019){Gim}, {Yun}, {Owen}, {Momjian}, {Miller},
  {Giavalisco}, {Wilson}, {Lowenthal}, {Aretxaga}, {Hughes}, {Morrison}, \&
  {Kawabe}}]{gim19}
{Gim}, H.~B., {Yun}, M.~S., {Owen}, F.~N., {et~al.} 2019, \apj, 875, 80,
  \dodoi{10.3847/1538-4357/ab1011}

\bibitem[{Gonz{\'a}lez-L{\'o}pez {et~al.}(2019)Gonz{\'a}lez-L{\'o}pez, Decarli,
  Pavesi, Walter, Aravena, Carilli, Boogaard, Popping, Weiss, Assef, Bauer,
  Bertoldi, Bouwens, Contini, Cortes, Cox, da~Cunha, Daddi, D{\'\i}az-Santos,
  Inami, Hodge, Ivison, Le~F{\`e}vre, Magnelli, Oesch, Riechers, Rix, Smail,
  Swinbank, Somerville, Uzgil, \& van~der Werf}]{gonzalez-lopez19}
Gonz{\'a}lez-L{\'o}pez, J., Decarli, R., Pavesi, R., {et~al.} 2019, The
  Astrophysical Journal, 882, 139, \dodoi{10.3847/1538-4357/ab3105}

\bibitem[{{Grogin} {et~al.}(2011){Grogin}, {Kocevski}, {Faber}, {Ferguson},
  {Koekemoer}, {Riess}, {Acquaviva}, {Alexander}, {Almaini}, {Ashby}, {Barden},
  {Bell}, {Bournaud}, {Brown}, {Caputi}, {Casertano}, {Cassata}, {Castellano},
  {Challis}, {Chary}, {Cheung}, {Cirasuolo}, {Conselice}, {Roshan Cooray},
  {Croton}, {Daddi}, {Dahlen}, {Dav{\'e}}, {de Mello}, {Dekel}, {Dickinson},
  {Dolch}, {Donley}, {Dunlop}, {Dutton}, {Elbaz}, {Fazio}, {Filippenko},
  {Finkelstein}, {Fontana}, {Gardner}, {Garnavich}, {Gawiser}, {Giavalisco},
  {Grazian}, {Guo}, {Hathi}, {H{\"a}ussler}, {Hopkins}, {Huang}, {Huang},
  {Jha}, {Kartaltepe}, {Kirshner}, {Koo}, {Lai}, {Lee}, {Li}, {Lotz}, {Lucas},
  {Madau}, {McCarthy}, {McGrath}, {McIntosh}, {McLure}, {Mobasher},
  {Moustakas}, {Mozena}, {Nandra}, {Newman}, {Niemi}, {Noeske}, {Papovich},
  {Pentericci}, {Pope}, {Primack}, {Rajan}, {Ravindranath}, {Reddy}, {Renzini},
  {Rix}, {Robaina}, {Rodney}, {Rosario}, {Rosati}, {Salimbeni}, {Scarlata},
  {Siana}, {Simard}, {Smidt}, {Somerville}, {Spinrad}, {Straughn}, {Strolger},
  {Telford}, {Teplitz}, {Trump}, {van der Wel}, {Villforth}, {Wechsler},
  {Weiner}, {Wiklind}, {Wild}, {Wilson}, {Wuyts}, {Yan}, \& {Yun}}]{grogin11}
{Grogin}, N.~A., {Kocevski}, D.~D., {Faber}, S.~M., {et~al.} 2011, \apjs, 197,
  35, \dodoi{10.1088/0067-0049/197/2/35}

\bibitem[{Guidetti {et~al.}(2017)Guidetti, Bondi, Prandoni, Muxlow, Beswick,
  Wrigley, Smail, McHardy, Thomson, Radcliffe, \& Argo}]{guidetti17}
Guidetti, D., Bondi, M., Prandoni, I., {et~al.} 2017, MNRAS, 471, 210,
  \dodoi{10.1093/mnras/stx1162}

\bibitem[{{Hale} {et~al.}(2021){Hale}, {McConnell}, {Thomson}, {Lenc}, {Heald},
  {Hotan}, {Leung}, {Moss}, {Murphy}, {Pritchard}, {Sadler}, {Stewart}, \&
  {Whiting}}]{hale21}
{Hale}, C.~L., {McConnell}, D., {Thomson}, A.~J.~M., {et~al.} 2021, \pasa, 38,
  e058, \dodoi{10.1017/pasa.2021.47}

\bibitem[{{Hale} {et~al.}(2023){Hale}, {Whittam}, {Jarvis}, {Best}, {Thomas},
  {Heywood}, {Prescott}, {Adams}, {Afonso}, {An}, {Bowler}, {Collier}, {Cook},
  {Dav{\'e}}, {Frank}, {Glowacki}, {Hatfield}, {Kolwa}, {Lovell}, {Maddox},
  {Marchetti}, {Morabito}, {Murphy}, {Prandoni}, {Randriamanakoto}, \&
  {Taylor}}]{hale23}
{Hale}, C.~L., {Whittam}, I.~H., {Jarvis}, M.~J., {et~al.} 2023, \mnras, 520,
  2668, \dodoi{10.1093/mnras/stac3320}

\bibitem[{{Heywood} {et~al.}(2021){Heywood}, {Murphy}, {Jim{\'e}nez-Andrade},
  {Armus}, {Cotton}, {DeCoursey}, {Dickinson}, {Lazio}, {Momjian}, {Penner},
  {Smail}, \& {Smirnov}}]{heywood21}
{Heywood}, I., {Murphy}, E.~J., {Jim{\'e}nez-Andrade}, E.~F., {et~al.} 2021,
  \apj, 910, 105, \dodoi{10.3847/1538-4357/abdf61}

\bibitem[{Huynh {et~al.}(2019)Huynh, Seymour, Norris, \& Galvin}]{huynh19}
Huynh, M.~T., Seymour, N., Norris, R.~P., \& Galvin, T. 2019, Monthly Notices
  of the Royal Astronomical Society, 491, 3395, \dodoi{10.1093/mnras/stz3187}

\bibitem[{{Ibar} {et~al.}(2009){Ibar}, {Ivison}, {Biggs}, {Lal}, {Best}, \&
  {Green}}]{ibar09}
{Ibar}, E., {Ivison}, R.~J., {Biggs}, A.~D., {et~al.} 2009, \mnras, 397, 281,
  \dodoi{10.1111/j.1365-2966.2009.14866.x}

\bibitem[{{Jim{\'e}nez-Andrade} {et~al.}(2019){Jim{\'e}nez-Andrade}, {Magnelli,
  B.}, {Karim, A.}, {Zamorani, G.}, {Bondi, M.}, {Schinnerer, E.}, {Sargent,
  M.}, {Romano-D\'{\i}az, E.}, {Novak, M.}, {Lang, P.}, {Bertoldi, F.},
  {Vardoulaki, E.}, {Toft, S.}, {Smolci\'{}c, V.}, {Harrington, K.}, {Leslie,
  S.}, {Delhaize, J.}, {Liu, D.}, {Karoumpis, C.}, {Kartaltepe, J.}, \&
  {Koekemoer, A. M.}}]{jimenezandrade19}
{Jim{\'e}nez-Andrade}, E.~F., {Magnelli, B.}, {Karim, A.}, {et~al.} 2019, A\&A,
  625, A114, \dodoi{10.1051/0004-6361/201935178}

\bibitem[{{Jim{\'e}nez-Andrade} {et~al.}(2021){Jim{\'e}nez-Andrade}, {Murphy},
  {Heywood}, {Smail}, {Penner}, {Momjian}, {Dickinson}, {Armus}, \&
  {Lazio}}]{jimenezandrade21}
{Jim{\'e}nez-Andrade}, E.~F., {Murphy}, E.~J., {Heywood}, I., {et~al.} 2021,
  \apj, 910, 106, \dodoi{10.3847/1538-4357/abe876}

\bibitem[{{Klein} {et~al.}(2018){Klein}, {Lisenfeld}, \& {Verley}}]{klein18}
{Klein}, U., {Lisenfeld}, U., \& {Verley}, S. 2018, \aap, 611, A55,
  \dodoi{10.1051/0004-6361/201731673}

\bibitem[{{Koekemoer} {et~al.}(2011){Koekemoer}, {Faber}, {Ferguson}, {Grogin},
  {Kocevski}, {Koo}, {Lai}, {Lotz}, {Lucas}, {McGrath}, {Ogaz}, {Rajan},
  {Riess}, {Rodney}, {Strolger}, {Casertano}, {Castellano}, {Dahlen},
  {Dickinson}, {Dolch}, {Fontana}, {Giavalisco}, {Grazian}, {Guo}, {Hathi},
  {Huang}, {van der Wel}, {Yan}, {Acquaviva}, {Alexander}, {Almaini}, {Ashby},
  {Barden}, {Bell}, {Bournaud}, {Brown}, {Caputi}, {Cassata}, {Challis},
  {Chary}, {Cheung}, {Cirasuolo}, {Conselice}, {Roshan Cooray}, {Croton},
  {Daddi}, {Dav{\'e}}, {de Mello}, {de Ravel}, {Dekel}, {Donley}, {Dunlop},
  {Dutton}, {Elbaz}, {Fazio}, {Filippenko}, {Finkelstein}, {Frazer}, {Gardner},
  {Garnavich}, {Gawiser}, {Gruetzbauch}, {Hartley}, {H{\"a}ussler},
  {Herrington}, {Hopkins}, {Huang}, {Jha}, {Johnson}, {Kartaltepe},
  {Khostovan}, {Kirshner}, {Lani}, {Lee}, {Li}, {Madau}, {McCarthy},
  {McIntosh}, {McLure}, {McPartland}, {Mobasher}, {Moreira}, {Mortlock},
  {Moustakas}, {Mozena}, {Nandra}, {Newman}, {Nielsen}, {Niemi}, {Noeske},
  {Papovich}, {Pentericci}, {Pope}, {Primack}, {Ravindranath}, {Reddy},
  {Renzini}, {Rix}, {Robaina}, {Rosario}, {Rosati}, {Salimbeni}, {Scarlata},
  {Siana}, {Simard}, {Smidt}, {Snyder}, {Somerville}, {Spinrad}, {Straughn},
  {Telford}, {Teplitz}, {Trump}, {Vargas}, {Villforth}, {Wagner}, {Wandro},
  {Wechsler}, {Weiner}, {Wiklind}, {Wild}, {Wilson}, {Wuyts}, \&
  {Yun}}]{koekemoer11}
{Koekemoer}, A.~M., {Faber}, S.~M., {Ferguson}, H.~C., {et~al.} 2011, \apjs,
  197, 36, \dodoi{10.1088/0067-0049/197/2/36}

\bibitem[{{Latif} {et~al.}(2024){Latif}, {Aftab}, \& {Whalen}}]{latif24}
{Latif}, M.~A., {Aftab}, A., \& {Whalen}, D.~J. 2024, arXiv e-prints,
  arXiv:2401.07910, \dodoi{10.48550/arXiv.2401.07910}

\bibitem[{{Leslie} {et~al.}(2020){Leslie}, {Schinnerer}, {Liu}, {Magnelli},
  {Algera}, {Karim}, {Davidzon}, {Gozaliasl}, {Jim{\'e}nez-Andrade}, {Lang},
  {Sargent}, {Novak}, {Groves}, {Smol{\v{c}}i{\'c}}, {Zamorani}, {Vaccari},
  {Battisti}, {Vardoulaki}, {Peng}, \& {Kartaltepe}}]{leslie20}
{Leslie}, S.~K., {Schinnerer}, E., {Liu}, D., {et~al.} 2020, \apj, 899, 58,
  \dodoi{10.3847/1538-4357/aba044}

\bibitem[{{Lindroos} {et~al.}(2016){Lindroos}, {Knudsen}, {Fan}, {Conway},
  {Coppin}, {Decarli}, {Drouart}, {Hodge}, {Karim}, {Simpson}, \&
  {Wardlow}}]{lindroos16}
{Lindroos}, L., {Knudsen}, K.~K., {Fan}, L., {et~al.} 2016, \mnras, 462, 1192,
  \dodoi{10.1093/mnras/stw1628}

\bibitem[{{Mancuso} {et~al.}(2015){Mancuso}, {Lapi}, {Cai}, {Negrello}, {De
  Zotti}, {Bressan}, {Bonato}, {Perrotta}, \& {Danese}}]{mancuso_15}
{Mancuso}, C., {Lapi}, A., {Cai}, Z.-Y., {et~al.} 2015, \apj, 810, 72,
  \dodoi{10.1088/0004-637X/810/1/72}

\bibitem[{{Mancuso} {et~al.}(2017){Mancuso}, {Lapi}, {Prandoni}, {Obi},
  {Gonzalez-Nuevo}, {Perrotta}, {Bressan}, {Celotti}, \& {Danese}}]{mancuso17}
{Mancuso}, C., {Lapi}, A., {Prandoni}, I., {et~al.} 2017, \apj, 842, 95,
  \dodoi{10.3847/1538-4357/aa745d}

\bibitem[{{Matthews} {et~al.}(2021{\natexlab{a}}){Matthews}, {Condon},
  {Cotton}, \& {Mauch}}]{matthews21}
{Matthews}, A.~M., {Condon}, J.~J., {Cotton}, W.~D., \& {Mauch}, T.
  2021{\natexlab{a}}, \apj, 909, 193, \dodoi{10.3847/1538-4357/abdd37}

\bibitem[{{Matthews} {et~al.}(2021{\natexlab{b}}){Matthews}, {Condon},
  {Cotton}, \& {Mauch}}]{matthews21b}
---. 2021{\natexlab{b}}, \apj, 914, 126, \dodoi{10.3847/1538-4357/abfaf6}

\bibitem[{{McMullin} {et~al.}(2007){McMullin}, {Waters}, {Schiebel}, {Young},
  \& {Golap}}]{mcmullin07}
{McMullin}, J.~P., {Waters}, B., {Schiebel}, D., {Young}, W., \& {Golap}, K.
  2007, in Astronomical Society of the Pacific Conference Series, Vol. 376,
  Astronomical Data Analysis Software and Systems XVI, ed. R.~A. {Shaw},
  F.~{Hill}, \& D.~J. {Bell}, 127

\bibitem[{Miettinen {et~al.}(2015)Miettinen, Smol{\v c}i{\'c}, Novak, Aravena,
  Karim, Masters, Riechers, {Bussmann, R. S.}, {McCracken, H. J.}, {Ilbert,
  O.}, {Bertoldi, F.}, {Capak, P.}, {Feruglio, C.}, {Halliday, C.},
  {Kartaltepe, J. S.}, {Navarrete, F.}, {Salvato, M.}, {Sanders, D.},
  {Schinnerer, E.}, \& {Sheth, K.}}]{miettinen15}
Miettinen, O., Smol{\v c}i{\'c}, V., Novak, M., {et~al.} 2015, A\&A, 577, A29,
  \dodoi{10.1051/0004-6361/201425032}

\bibitem[{{Miley}(1980)}]{miley80}
{Miley}, G. 1980, \araa, 18, 165, \dodoi{10.1146/annurev.aa.18.090180.001121}

\bibitem[{Mohan \& Rafferty(2015)}]{Mohan15}
Mohan, N., \& Rafferty, D. 2015, Astrophysics Source Code Library, 1502.007

\bibitem[{{Morrison} {et~al.}(2010){Morrison}, {Owen}, {Dickinson}, {Ivison},
  \& {Ibar}}]{morrison10}
{Morrison}, G.~E., {Owen}, F.~N., {Dickinson}, M., {Ivison}, R.~J., \& {Ibar},
  E. 2010, \apjs, 188, 178, \dodoi{10.1088/0067-0049/188/1/178}

\bibitem[{Murphy {et~al.}(2018{\natexlab{a}})Murphy, J.~Condon, Alberdi,
  Barcos-Mu{\~n}ozarcos, J.~Beswick, Brinks, Dong, S.~Evans, E.~Johnson, Rober
  C.~Kennicutt, T.~Linden, W.~B.~Muxlow, Perez-Torres, Schinnerer, Sargent,
  Tabatabaei, \& L.~Turner}]{murphy18}
Murphy, E., J.~Condon, J., Alberdi, A., {et~al.} 2018{\natexlab{a}}, Science
  with an ngVLA: Radio Continuum Emission from Galaxies: An Accounting of
  Energetic Processes

\bibitem[{{Murphy}(2022)}]{murphy22}
{Murphy}, E.~J. 2022, Universe, 8, 329, \dodoi{10.3390/universe8060329}

\bibitem[{Murphy {et~al.}(2018{\natexlab{b}})Murphy, Linden, Dong, Hensley,
  Momjian, Helou, \& Evans}]{murphy18_ames}
Murphy, E.~J., Linden, S.~T., Dong, D., {et~al.} 2018{\natexlab{b}}, The
  Astrophysical Journal, 862, 20, \dodoi{10.3847/1538-4357/aac5f5}

\bibitem[{Murphy {et~al.}(2017)Murphy, Momjian, Condon, Chary, Dickinson,
  Inami, Taylor, \& Weiner}]{murphy17}
Murphy, E.~J., Momjian, E., Condon, J.~J., {et~al.} 2017, ApJ, 839, 35,
  \dodoi{10.3847/1538-4357/aa62fd}

\bibitem[{Murphy {et~al.}(2011)Murphy, Condon, Schinnerer, Kennicutt, Calzetti,
  Armus, Helou, Turner, Aniano, Beir{\~{a}}o, Bolatto, Brandl, Croxall, Dale,
  Meyer, Draine, Engelbracht, Hunt, Hao, Koda, Roussel, Skibba, \&
  Smith}]{murphy11}
Murphy, E.~J., Condon, J.~J., Schinnerer, E., {et~al.} 2011, ApJ, 737, 67,
  \dodoi{10.1088/0004-637x/737/2/67}

\bibitem[{Muxlow {et~al.}(2005)Muxlow, Richards, Garrington, Wilkinson,
  Anderson, Richards, Axon, Fomalont, Kellermann, Partridge, \&
  Windhorst}]{muxlow05}
Muxlow, T. W.~B., Richards, A. M.~S., Garrington, S.~T., {et~al.} 2005, MNRAS,
  358, 1159.
\newblock \url{http://dx.doi.org/10.1111/j.1365-2966.2005.08824.x}

\bibitem[{{Muxlow} {et~al.}(2020){Muxlow}, {Thomson}, {Radcliffe}, {Wrigley},
  {Beswick}, {Smail}, {McHardy}, {Garrington}, {Ivison}, {Jarvis}, {Prandoni},
  {Bondi}, {Guidetti}, {Argo}, {Bacon}, {Best}, {Biggs}, {Chapman}, {Coppin},
  {Chen}, {Garratt}, {Garrett}, {Ibar}, {Kneib}, {Knudsen}, {Koopmans},
  {Morabito}, {Murphy}, {Njeri}, {Pearson}, {P{\'e}rez-Torres}, {Richards},
  {R{\"o}ttgering}, {Sargent}, {Serjeant}, {Simpson}, {Simpson}, {Swinbank},
  {Varenius}, \& {Venturi}}]{muxlow20}
{Muxlow}, T.~W.~B., {Thomson}, A.~P., {Radcliffe}, J.~F., {et~al.} 2020,
  \mnras, 495, 1188, \dodoi{10.1093/mnras/staa1279}

\bibitem[{{Novak} {et~al.}(2017){Novak}, {Smol{\v c}i{\'c}, V.}, {Delhaize,
  J.}, {Delvecchio, I.}, {Zamorani, G.}, {Baran, N.}, {Bondi, M.}, {Capak, P.},
  {Carilli, C. L.}, {Ciliegi, P.}, {Civano, F.}, {Ilbert, O.}, {Karim, A.},
  {Laigle, C.}, {Le F{\`e}vre, O.}, {Marchesi, S.}, {McCracken, H.},
  {Miettinen, O.}, {Salvato, M.}, {Sargent, M.}, {Schinnerer, E.}, \& {Tasca,
  L.}}]{novak17}
{Novak}, M., {Smol{\v c}i{\'c}, V.}, {Delhaize, J.}, {et~al.} 2017, A\&A, 602,
  A5, \dodoi{10.1051/0004-6361/201629436}

\bibitem[{{Oesch} {et~al.}(2023){Oesch}, {Brammer}, {Naidu}, {Bouwens},
  {Chisholm}, {Illingworth}, {Matthee}, {Nelson}, {Qin}, {Reddy}, {Shapley},
  {Shivaei}, {van Dokkum}, {Weibel}, {Whitaker}, {Wuyts}, {Covelo-Paz},
  {Endsley}, {Fudamoto}, {Giovinazzo}, {Herard-Demanche}, {Kerutt},
  {Kramarenko}, {Labbe}, {Leonova}, {Lin}, {Magee}, {Marchesini}, {Maseda},
  {Mason}, {Matharu}, {Meyer}, {Neufeld}, {Prieto Lyon}, {Schaerer}, {Sharma},
  {Shuntov}, {Smit}, {Stefanon}, {Wyithe}, \& {Xiao}}]{oesch23}
{Oesch}, P.~A., {Brammer}, G., {Naidu}, R.~P., {et~al.} 2023, \mnras, 525,
  2864, \dodoi{10.1093/mnras/stad2411}

\bibitem[{{Owen}(2018)}]{owen18}
{Owen}, F.~N. 2018, \apjs, 235, 34, \dodoi{10.3847/1538-4365/aab4a1}

\bibitem[{{Radcliffe} {et~al.}(2018){Radcliffe}, {Garrett}, {Muxlow},
  {Beswick}, {Barthel}, {Deller}, {Keimpema}, {Campbell}, \&
  {Wrigley}}]{radcliffe18}
{Radcliffe}, J.~F., {Garrett}, M.~A., {Muxlow}, T.~W.~B., {et~al.} 2018, \aap,
  619, A48, \dodoi{10.1051/0004-6361/201833399}

\bibitem[{{Rau} \& {Cornwell}(2011)}]{rau-cornwell11}
{Rau}, U., \& {Cornwell}, T.~J. 2011, \aap, 532, A71,
  \dodoi{10.1051/0004-6361/201117104}

\bibitem[{{Richards} {et~al.}(1999){Richards}, {Fomalont}, {Kellermann},
  {Windhorst}, {Partridge}, {Cowie}, \& {Barger}}]{richards99}
{Richards}, E.~A., {Fomalont}, E.~B., {Kellermann}, K.~I., {et~al.} 1999, \apj,
  526, L73, \dodoi{10.1086/312373}

\bibitem[{{Richards} {et~al.}(1998){Richards}, {Kellermann}, {Fomalont},
  {Windhorst}, \& {Partridge}}]{richards98}
{Richards}, E.~A., {Kellermann}, K.~I., {Fomalont}, E.~B., {Windhorst}, R.~A.,
  \& {Partridge}, R.~B. 1998, \aj, 116, 1039, \dodoi{10.1086/300489}

\bibitem[{{Robitaille} \& {Bressert}(2012)}]{2012ascl.soft08017R}
{Robitaille}, T., \& {Bressert}, E. 2012, {APLpy: Astronomical Plotting Library
  in Python}, Astrophysics Source Code Library, record ascl:1208.017.
\newblock \doeprint{1208.017}

\bibitem[{Rujopakarn {et~al.}(2016)Rujopakarn, Dunlop, Rieke, Ivison, Cibinel,
  Nyland, Jagannathan, Silverman, Alexander, Biggs, Bhatnagar, Ballantyne,
  Dickinson, Elbaz, Geach, Hayward, Kirkpatrick, McLure, Micha{\l}owski,
  Miller, Narayanan, Owen, Pannella, Papovich, Pope, Rau, Robertson, Scott,
  Swinbank, van~der Werf, van Kampen, Weiner, \& Windhorst}]{rujopakarn16}
Rujopakarn, W., Dunlop, J.~S., Rieke, G.~H., {et~al.} 2016, ApJ, 833, 12,
  \dodoi{10.3847/0004-637X/833/1/12}

\bibitem[{{Sadler} {et~al.}(2008){Sadler}, {Ricci}, {Ekers}, {Sault},
  {Jackson}, \& {de Zotti}}]{sadler08}
{Sadler}, E.~M., {Ricci}, R., {Ekers}, R.~D., {et~al.} 2008, \mnras, 385, 1656,
  \dodoi{10.1111/j.1365-2966.2008.12955.x}

\bibitem[{{Sadler} {et~al.}(2006){Sadler}, {Ricci}, {Ekers}, {Ekers},
  {Hancock}, {Jackson}, {Kesteven}, {Murphy}, {Phillips}, {Reinfrank},
  {Staveley-Smith}, {Subrahmanyan}, {Walker}, {Wilson}, \& {de
  Zotti}}]{sadler06}
---. 2006, \mnras, 371, 898, \dodoi{10.1111/j.1365-2966.2006.10729.x}

\bibitem[{{Schinnerer} {et~al.}(2007){Schinnerer}, {Smol{\v{c}}i{\'c}},
  {Carilli}, {Bondi}, {Ciliegi}, {Jahnke}, {Scoville}, {Aussel}, {Bertoldi},
  {Blain}, {Impey}, {Koekemoer}, {Le Fevre}, \& {Urry}}]{schinnerer07}
{Schinnerer}, E., {Smol{\v{c}}i{\'c}}, V., {Carilli}, C.~L., {et~al.} 2007,
  \apjs, 172, 46, \dodoi{10.1086/516587}

\bibitem[{{Seymour} {et~al.}(2004){Seymour}, {McHardy}, \& {Gunn}}]{seymour04}
{Seymour}, N., {McHardy}, I.~M., \& {Gunn}, K.~F. 2004, \mnras, 352, 131,
  \dodoi{10.1111/j.1365-2966.2004.07904.x}

\bibitem[{{Simpson}(2017)}]{simpson17}
{Simpson}, C. 2017, Royal Society Open Science, 4, 170522,
  \dodoi{10.1098/rsos.170522}

\bibitem[{{Smol{\v c}i{\'c}} {et~al.}(2017{\natexlab{a}}){Smol{\v c}i{\'c}},
  Novak, Bondi, Ciliegi, Mooley, Schinnerer, {Zamorani, G.}, {Navarrete, F.},
  {Bourke, S.}, {Karim, A.}, {Vardoulaki, E.}, {Leslie, S.}, {Delhaize, J.},
  {Carilli, C. L.}, {Myers, S. T.}, {Baran, N.}, {Delvecchio, I.}, {Miettinen,
  O.}, {Banfield, J.}, {Balokovi{\'c}, M.}, {Bertoldi, F.}, {Capak, P.},
  {Frail, D. A.}, {Hallinan, G.}, {Hao, H.}, {Herrera Ruiz, N.}, {Horesh, A.},
  {Ilbert, O.}, {Intema, H.}, {Jeli{\'c}, V.}, {Kl{\"o}ckner, H.-R.}, {Krpan,
  J.}, {Kulkarni, S. R.}, {McCracken, H.}, {Laigle, C.}, {Middleberg, E.},
  {Murphy, E. J.}, {Sargent, M.}, {Scoville, N. Z.}, \& {Sheth,
  K.}}]{smolcic17}
{Smol{\v c}i{\'c}}, V., Novak, M., Bondi, M., {et~al.} 2017{\natexlab{a}},
  A\&A, 602, A1, \dodoi{10.1051/0004-6361/201628704}

\bibitem[{{Smol{\v c}i{\'c}} {et~al.}(2017{\natexlab{b}}){Smol{\v c}i{\'c}},
  {Delvecchio, I.}, {Zamorani, G.}, {Baran, N.}, {Novak, M.}, {Delhaize, J.},
  {Schinnerer, E.}, {Berta, S.}, {Bondi, M.}, {Ciliegi, P.}, {Capak, P.},
  {Civano, F.}, {Karim, A.}, {Le Fevre, O.}, {Ilbert, O.}, {Laigle, C.},
  {Marchesi, S.}, {McCracken, H. J.}, {Tasca, L.}, {Salvato, M.}, \&
  {Vardoulaki, E.}}]{smolcic17b}
{Smol{\v c}i{\'c}}, V., {Delvecchio, I.}, {Zamorani, G.}, {et~al.}
  2017{\natexlab{b}}, A\&A, 602, A2, \dodoi{10.1051/0004-6361/201630223}

\bibitem[{{Somerville} {et~al.}(2004){Somerville}, {Lee}, {Ferguson},
  {Gardner}, {Moustakas}, \& {Giavalisco}}]{somerville04}
{Somerville}, R.~S., {Lee}, K., {Ferguson}, H.~C., {et~al.} 2004, \apjl, 600,
  L171, \dodoi{10.1086/378628}

\bibitem[{{Tabatabaei} {et~al.}(2017){Tabatabaei}, {Schinnerer}, {Krause},
  {Dumas}, {Meidt}, {Damas-Segovia}, {Beck}, {Murphy}, {Mulcahy}, {Groves},
  {Bolatto}, {Dale}, {Galametz}, {Sandstrom}, {Boquien}, {Calzetti},
  {Kennicutt}, {Hunt}, {De Looze}, \& {Pellegrini}}]{tabatabaei17}
{Tabatabaei}, F.~S., {Schinnerer}, E., {Krause}, M., {et~al.} 2017, \apj, 836,
  185, \dodoi{10.3847/1538-4357/836/2/185}

\bibitem[{{Tadhunter}(2016)}]{tadhunter16}
{Tadhunter}, C. 2016, \aapr, 24, 10, \dodoi{10.1007/s00159-016-0094-x}

\bibitem[{Thomson {et~al.}(2019)Thomson, Smail, Swinbank, Simpson, Arumugam,
  Stach, Murphy, Rujopakarn, Almaini, An, Blain, Chen, Cooke, Dudzevi{\v
  c}i{\=u}t{\.e}, Edge, Farrah, Gullberg, Hartley, Ibar, Maltby,
  Micha{\l}owski, Simpson, van~der Werf, \& Wardlow}]{thomson_19}
Thomson, A.~P., Smail, I., Swinbank, A.~M., {et~al.} 2019, The Astrophysical
  Journal, 883, 204, \dodoi{10.3847/1538-4357/ab32e7}

\bibitem[{{Tisani{\'c}, K.} {et~al.}(2020){Tisani{\'c}, K.}, {Smol{\v c}i{\'c},
  V.}, {Imbri{\v s}ak, M.}, {Bondi, M.}, {Zamorani, G.}, {Ceraj, L.},
  {Vardoulaki, E.}, \& {Delhaize, J.}}]{tisanic20}
{Tisani{\'c}, K.}, {Smol{\v c}i{\'c}, V.}, {Imbri{\v s}ak, M.}, {et~al.} 2020,
  A\&A, 643, A51, \dodoi{10.1051/0004-6361/201937114}

\bibitem[{{van der Vlugt} {et~al.}(2021){van der Vlugt}, {Algera}, {Hodge},
  {Novak}, {Radcliffe}, {Riechers}, {R{\"o}ttgering}, {Smol{\v{c}}i{\'c}}, \&
  {Walter}}]{vandervlugt21}
{van der Vlugt}, D., {Algera}, H.~S.~B., {Hodge}, J.~A., {et~al.} 2021, \apj,
  907, 5, \dodoi{10.3847/1538-4357/abcaa3}

\bibitem[{Vardoulaki {et~al.}(2019)Vardoulaki, Jim{\'e}nez~Andrade, Karim,
  Novak, Leslie, {et~al.}}]{vardoulaki19}
Vardoulaki, E., Jim{\'e}nez~Andrade, E.~F., Karim, A., {et~al.} 2019, A\&A in
  press. arXiv:1901.10168

\bibitem[{{White} {et~al.}(2012){White}, {Hatsukade}, {Pearson}, {Takagi},
  {Sedgwick}, {Matsuura}, {Matsuhara}, {Serjeant}, {Nakagawa}, {Lee}, {Oyabu},
  {Jeong}, {Shirahata}, {Kohno}, {Yamamura}, {Hanami}, {Goto}, {Makiuti},
  {Clements}, {Malek}, \& {Khan}}]{white12}
{White}, G.~J., {Hatsukade}, B., {Pearson}, C., {et~al.} 2012, \mnras, 427,
  1830, \dodoi{10.1111/j.1365-2966.2012.21684.x}

\bibitem[{{Whittam} {et~al.}(2016){Whittam}, {Riley}, {Green}, {Davies},
  {Franzen}, {Rumsey}, {Schammel}, \& {Waldram}}]{whittam16}
{Whittam}, I.~H., {Riley}, J.~M., {Green}, D.~A., {et~al.} 2016, \mnras, 457,
  1496, \dodoi{10.1093/mnras/stv2960}

\bibitem[{Windhorst {et~al.}(1990)Windhorst, Mathis, \&
  Neuschaefer}]{windhorst90}
Windhorst, R.~A., Mathis, D., \& Neuschaefer, L. 1990, in ASP Conf. Ser.,
  Evolution of the universe of galaxies, 389

\end{thebibliography}
\bibliographystyle{aasjournal}



\end{document}